\documentclass[prc,aps,onecolumn,showpcs,amsmath,amssymb,nofootinbib,notitlepage,preprintnumbers,balancelastpage,longbibliography]{revtex4-2}

\usepackage{amsfonts}
\usepackage{amsmath}
\usepackage{hyperref}
\hypersetup{breaklinks=true, colorlinks=true, citecolor=blue, linkcolor=blue, urlcolor=blue,filecolor=blue}
\usepackage{anyfontsize}
\usepackage{ulem}

\usepackage{lineno}
\usepackage{cancel}
\usepackage[utf8]{inputenc}

\usepackage{booktabs}
\usepackage{graphicx}

\usepackage{array}
\usepackage{siunitx}
\usepackage{makecell}
\usepackage{rotating}

\DeclareSIUnit\ppb{ppb}

\usepackage{multirow}
\usepackage{mathrsfs}
\usepackage{enumitem}
\usepackage{feynmp}
\usepackage{amsmath,amssymb,amsthm,amsfonts}
\usepackage{subfigure}
\usepackage{dcolumn}	
\usepackage{bm}		
\usepackage{epsfig}
\usepackage{epstopdf}
\usepackage{setspace}
\usepackage[usenames, dvipsnames]{color}
\usepackage{slashed}
\usepackage{comment}
\usepackage{enumitem}
\usepackage{siunitx}
\usepackage{multirow}
\usepackage{feynmp}
\usepackage{datetime}
\usepackage{comment}

\usepackage{enumitem}
\usepackage{titlesec, titletoc}
\usepackage{latexsym}

\newcommand{\be}{\begin{equation}\begin{aligned}}
\newcommand{\ee}{\end{aligned}\end{equation}}
\newcommand{\beq}{\begin{equation}}
\newcommand{\eeq}{\end{equation}}
\newcommand{\beqa}{\begin{eqnarray}}
\newcommand{\eeqa}{\end{eqnarray}}

\newcommand{\ME}{\mathcal{M}}
\DeclareMathAlphabet{\mathpzc}{OT1}{pzc}{m}{it}

\begin{document}

\title{\texorpdfstring{Quantum Chromodynamics Resolution of the ATOMKI Anomaly in ${\rm ^4He}$ Nuclear Transitions}
{Quantum Chromodynamics Resolution of the ATOMKI Anomaly in 4He Nuclear Transitions}}

\author{Valery~Kubarovsky}
\email{vpk@jlab.org}
\affiliation{Thomas Jefferson National Accelerator Laboratory, Newport News, VA 23606, USA}
\author{Jennifer~Rittenhouse~West}
\email{rittenhouse@berkeley.edu}
\affiliation{Lawrence Berkeley National Laboratory, Berkeley, CA 94720, USA}
\affiliation{EIC Center at Thomas Jefferson National Accelerator Laboratory, Newport News, VA 23606, USA}
\author{Stanley~J.~Brodsky}
\email{sjbth@slac.stanford.edu}
\affiliation{SLAC National Accelerator Laboratory, Stanford University, Stanford, CA 94309, USA}

\begin{abstract}
Observations of anomalous angular correlations in electron-positron pairs produced from excited states of $^{4}$He, $^{8}$Be and $^{12}$C nuclei have been suggested as due to the creation and subsequent decay of a new light particle of mass $\sim$17 MeV.  In this work, we investigate the possibility that the source of the observed signals is a set of new excitation channels created by the 12-quark hidden-color Fock state within the ${\rm {^4He}}$ nuclear wavefunction dubbed the ``hexadiquark.''  We calculate the invariant $e^+e^-$ mass spectrum for the electromagnetic transition from a new excitation of $^{4}$He, estimating its differential and total decay width. We find that we can fit the shape of the anomalous signal with the QCD Fock state at excitation energy $\rm E^{*} = 17.9 \pm 1$ MeV and a Gaussian form factor for the electromagnetic decay.  We address the physical issues with the fit parameters using properties of the hexadiquark state, in particular the three weakly repulsive $\rm 6_C$ interactions of $\rm SU(3)_C$ between diquark pairs.  Experimental tests of our model are described in detail.  In light of this work, we emphasize the need for independent experimental confirmation or refutation of the ATOMKI results as well as dedicated experiments to search for the proposed new excitations of ${\rm ^4He}$ and other $\alpha$-cluster nuclei.
\end{abstract}

\maketitle

\section{Introduction}
\label{sec:intro}

Observations by the ATOMKI collaboration of anomalous angular correlations in  electron-positron pairs produced 
 in the nuclear decays $ {\rm ^4He}^*\to {\rm ^4He}+ e^+e^-$ 
\cite{Krasznahorkay:2021joi,Krasznahorkay:2019lyl}, ${\rm ^8Be}^*\to {\rm ^8Be}+ e^+e^-$  \cite{Krasznahorkay:2015iga} and $ {\rm ^{12}C}^*\to {\rm ^{12}C}+ e^+e^-$ \cite{Krasznahorkay:2022pxs} have been attributed to the creation and subsequent decay of a new light particle of mass of $\sim 17$ MeV, dubbed the X17.  
The same group has reported observations of the lepton pair in the off-resonance region of $\rm {^7Li}(p,e^+e^-){^8Be}$ direct proton-capture reactions \cite{Sas:2022pgm}.  The signals have generated a great deal of theoretical interest in both the particle and nuclear physics communities \cite{Battaglieri:2017aum,Backens:2021qkv,Fornal:2017msy,Feng:2016ysn,Viviani:2021stx,Agrawal:2021dbo,Alexander:2016aln,Kozaczuk:2016nma,Kitahara:2016zyb,Feng:2020mbt,Zhang:2017zap,Zhang:2020ukq,Frugiuele:2016rii,NA64:2018lsq,Castro:2021gdf,Dror:2017ehi,Alves:2017avw,FASER:2018eoc,DelleRose:2018pgm,Hayes:2021hin,NA64:2020xxh,Seto:2016pks,Seto:2020jal,Ahmidouch:2021edo}.  In this work, we focus on the $\rm ^4He$ experiment in which the observed invariant mass $m_{e^+e^-}$ of the lepton pair was found to be $m_{\mathrm{X}}=16.94 \pm 0.12(\text{stat.}) \pm 0.21 \text { (syst.) } \mathrm{MeV}$.  We show that excitation and decay of the recently proposed QCD hidden-color Fock state within the $^4\rm He$ nuclear wavefunction \cite{West:2020rlk} provides a viable and compelling explanation of the ATOMKI phenomenon as well as a clear set of experimental predictions.  The model acts in all A$\geq$4 nuclei and explains why the X17 signal has only been observed in $\alpha$-nuclei.

The light-front Fock state expansion of QCD has led to new perspectives for the nonperturbative eigenstructure
of hadrons. The $[ud]$ scalar diquark in the $\bar{3}_C$ representation of $\rm SU(3)_C$, for example, is a configuration that appears to play a fundamental role in hadron spectroscopy such as the $|u [ud]\rangle$ Fock state in the proton wavefunction and baryonic Regge trajectories \cite{Dosch:2015nwa,deTeramond:2014asa}. Tetraquarks are very likely bound states of diquarks and antidiquarks \cite{tHooft:2008rus,Masjuan:2017fzu,Anselmino:1992vg,Nielsen:2018uyn}.  Quark-antiquark, quark-diquark, and diquark-antidiquark bound states all have an identical $3_C - \bar{3}_C$ color-confining interaction and therefore matching spectroscopy in QCD \cite{Brodsky:2014yha}.  

The 12-quark (6-diquark) hidden-color state $|[ud][ud][ud][ud][ud][ud]\rangle$ dubbed the ``hexadiquark'' is a novel color-singlet of QCD \cite{West:2020rlk}.  Such hidden-color states are predictions of the $\rm SU(3)_C$ group theory basis of QCD and have been studied for over four decades \cite{Brodsky:1976rz,Harvey:1980rva,Brodsky:1983vf,Harvey:1981udr,Bashkanov:2013cla,Miller:2013hla}. In general, a hadronic or nuclear eigenstate of the QCD Hamiltonian is a sum over all color-singlet Fock states which match its quantum numbers.  This has a special impact for the $^4\rm He$ nucleus, since the hexadiquark Fock state is an unusually low mass state with the same quantum numbers as the nuclear Fock state $|nnpp\rangle$.  Nuclear wavefunctions are typically dominated by the neutron and proton Fock state, i.e., the multiple (A) 3-quark color-singlet Fock state.  In the $^4\rm He$ nuclear wavefunction, this is the $|nnpp\rangle$ state containing A=4 color singlets (2 protons and 2 neutrons).  In addition to this state, every 12-quark combination with the same quantum numbers of $^4\rm He$ is also in the wavefunction - including the single color-singlet hexadiquark.  The hexadiquark is a special hidden-color state because it is an unusually low mass combination of quarks - all spin-0, isospin-0 and S-wave combinations - that obeys the spin-statistics theorem at every stage of the build.  This means it has the opportunity to have a larger effect on nuclear physics than is typical.  The six scalar diquarks within the hexadiquark hidden-color state have predicted experimental signatures in diffractive dissociation of $^4 \rm He$ nuclei, as recently published \cite{RittenhouseWest:2019sar}.

QCD also predicts orbital and radial excitations between the $[ud]$ diquark components of the hexadiquark and therefore new excitations of $\rm ^4He$ beyond the standard excitations predicted by nucleonic degrees of freedom.  The excitation energy of the hexadiquark can be below the energy required to produce nuclear decays such as $\rm ^4He \rightarrow p + \rm ^3H$ and have evaded detection thus far while containing new experimental predictions as will be demonstrated. The first hints of an excited state of $^4$He were discovered over 80 years ago \cite{Jarmie:1959zz,Jarmie:1963zz} and the question of how a new subdominant excited state could have been missed must be addressed. Hidden-color states cannot decay through conventional nuclear channels as all of the constituents have color charge (specifically quarks, diquarks and duo-diquarks).  The weakly repulsive QCD bonds between diquarks, utilizing the symmetric sextet $\rm 6_C$ channel of $\rm SU(3)_C$, allow for low (nuclear) energy excitations and require dedicated searches for their signatures.  An exhaustive list of QCD decay channels and experimental signatures is presented in Section \ref{sec:hdq}.

\section{Fitting the nuclear decay amplitude for \texorpdfstring{${\rm ^4He}^*\rightarrow {\rm ^4He}+ e^+e^-$}{excited 4He to ground state 4He plus an electron-positron pair}}

\begin{figure}[ht]
\begin{center}
\includegraphics[width=0.5\textwidth]{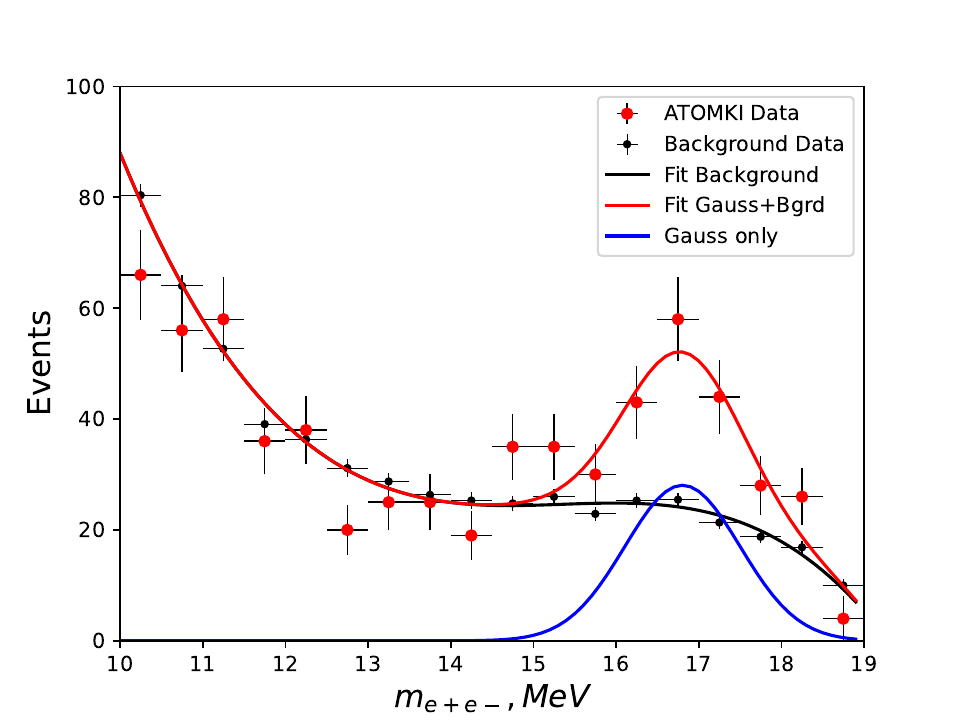}
\vspace*{-0.0cm}
\caption{Fit to the ATOMKI data (in red) by a Gaussian function and background distribution (in black) of the $m_{e^+e^-}$ invariant mass distribution based on \cite{Krasznahorkay:2019lyl} and cross-checked with \cite{NA64:2020xxh}.  The mean value of the fitted mass is $\sim 17$ MeV with width $\sigma \sim 0.7$ MeV.  The signal alone is shown in blue.}
\label{fig:fig_gauss0}
\end{center}
\end{figure}

The signal
template we use to match the ATOMKI $e^+e^-$ angular correlation is a narrow Gaussian peak in the electron-positron invariant mass with a mean value $\sim 17$ MeV and width $\sigma \sim 0.7$ MeV (see Figure~\ref{fig:fig_gauss0}) based on unpublished work by the ATOMKI collaboration \cite{Krasznahorkay:2019lyl} and cross-checked with the NA64 X17 invariant mass simulation \cite{NA64:2020xxh}. We calculate the invariant mass $m_{e^+e^-}$ spectrum in the electromagnetic transition of an excited state of $\rm ^4He$,  ${\rm ^4He}^*({\rm E^*)}\to {\rm ^4He}+ e^+e^-$, where $\rm E^*$ is the excited state energy.  The mass difference between the first known excited state and the ground state of the helium nucleus is $20.21$ MeV \cite{Walcher:1970vkv} and $\rm E^*$ is initially fixed to this value. Its decay to the $\rm ^4He$ ground state via a virtual photon cannot explain the narrow width of the observed X17 signal.  The value of $\rm E^*$ is subsequently allowed to float to a different value corresponding to the proposed excitation of the 12-quark color-singlet hexadiquark Fock state in the $\rm ^4He$ nuclear wavefunction \cite{West:2020rlk}.

The invariant mass distribution of the lepton
pair $m_{e^+e^-}$ from the decay of excited states of $\rm ^4He$  has not been calculated until now. The result of our calculation shows that decay due to an excitation of the hexadiquark Fock state in the eigensolution of $\rm ^4He$ can describe the signal and its narrow width;   unconventional signals are to be expected from the existence of the new hidden-color QCD state.

We begin with the square of the amplitude for the electromagnetic decay  ${\rm ^4He^*}  \to   {\rm ^4He}+e^+e^-$, given by \cite{Feng:2020mbt}

\beq
\left|\mathcal{M}\right|^2=\sum_\text{spin}\left|{M} \right|^2=\frac{e^4 \,C_{\text{E0}}^2}{\Lambda^4}\left[2(p_+\cdot p_0)(p_-\cdot p_0)-m_{0}^2(p_+\cdot p_-)-m_e^2m_{0}^2 \right] ,
\label{Eq1}
\eeq
\noindent
where $p_+$ and $p_-$ are the 4-momenta of the positron and electron respectively, 
 $p_0$ and $m_0$ are the 4-momentum and mass of the ground state of $\rm ^4He$ and $m_e$ is the electron mass.  Final state spins have been summed over. The definitions of $\Lambda$ as the nuclear energy scale and $ C^2_{E0}$ as the Wilson coefficient of the decay operator are contained in \cite{Feng:2020mbt}. These constants are not relevant for the analysis carried out in this work as they affect only the normalization which we match to the signal.

 The matrix element depends on only two kinematical variables, $q^2$ and $\cos{\theta^*}$, where 
 $q=p_++p_-$ is the 4-momentum of the virtual photon, $q^2=m^2_{e^+e^-}$ is  the $e^+e^-$ invariant mass squared 
 and $\theta^*$ is the electron angle in the virtual photon center of mass system.  Rewriting Eq.\ref{Eq1} in terms of these variables gives
 \begin{equation} \label{eq4}
 \begin{split}
\left| \ME (q^2,\cos{\theta^*})\right|^2 = 
\frac{\alpha^2 C^2_{E0}}{\Lambda^4}
 \frac{(4\pi)^2}{ 8q^2 } \left[ q^2-\cos^2{\theta^*} ( q^2-4 m_e^2)\right] \left[ m^4+(m^2_0-q^2)^2-2 m(m_0^2+q^2) \right]
\end{split}
\end{equation}
 
\noindent
 where $m$ is the mass of the excited ${\rm ^4He^*}$ state.

The differential width $d\Gamma$ for the three body decay is given by

\begin{equation} \label{eq7}
\begin{split}
d\Gamma=\frac{1}{(2\pi)^5}\frac{1}{16 m^2}\left|\ME\right|^2 |p_e^*| |p_{0}| dm_{e^+e^-} d\Omega^*_e d\Omega_{0}
\end{split}
\end{equation}
\noindent
where 
$d\Omega=d\phi~ d\cos{\theta}$, 
$(\left|p^*_e\right|, \Omega^*_e)$ is the electron momentum in the center of mass of the electron-positron pair and 
$(\left|p_0\right|, \Omega_0)$ is  the ${\rm ^4He}$ momentum in the rest frame of the ${\rm ^4He}^*.$  The minimum value of the $m_{e^+e^-}$ invariant mass is $2m_e$ and the maximum value is determined by the mass difference between the excited and ground ${\rm ^4He}$ states, $\Delta m=$20.21 MeV.

The integration of the matrix element $\left| \ME (q^2,\cos{\theta^*})\right|^2$ over angles $d\Omega^*_e\Omega_0$ gives
\begin{equation}\label{ME_full}
\left| \ME_q (q^2)\right|^2=\int \left| \ME (q^2,\cos{\theta^*})\right|^2 d\Omega^*_ed\Omega_0 = \frac{\alpha^2 C^2_{E0}}{\Lambda^4} \frac{(4\pi)^4}{12q^2} (q^2+2 m^2_e) \left[ m^4 + (m_0^2 - q^2)^2 - 2m^2 (m_0^2 + q^2) \right] .
\end{equation}

The matrix element $\left| \ME_q(q^2) \right|^2$ simplifies in the limit $m_e=0$, to 
\begin{equation} \label{ME_me}
\begin{split}
\left| \ME_q(q^2) \right|^2 = \frac{\alpha^2 C^2_{E0}}{\Lambda^4} \frac{(4\pi)^4}{12}  \left[ m^4 + (m_0^2 - q^2)^2 - 2m^2 (m_0^2 + q^2)  \right].
\end{split}
\end{equation}

There is only one variable left, the invariant electron-positron mass $m_{e^+e^-}=\sqrt{q^2}=\sqrt{(p_++p_-)^2}$, and we find 
\begin{equation} \label{eq8}
\begin{split}
d\Gamma=\frac{1}{(2\pi)^5}\frac{1}{16 m^2}\left|\ME_q\right|^2 |p_e^*| |p_{0}| dm_{e^+e^-}.
\end{split}
\end{equation}
\noindent

The numerical integration over the three-body phase space is 
\begin{equation}\label{FW_2}
\Gamma_{E0}=\int_{2m_e}^{m-m_0}dm_{e^+e^-}\left| \ME_q (q^2)\right|^2 |p_e^*| |p_{He}|=
 \frac{\alpha^2 C^2_{E0}}{\Lambda^4}~\frac {\Delta m^5}{60\pi}0.982,
 \end{equation}
which coincides with the approximate solution (Eq.~\ref{FW_me}, below) to an accuracy of $2\%$.

In the approximation $q^2/m^2< \Delta m^2/m^2= 3\cdot10^{-5}\ll 1$ and negligible electron mass, the differential width $d\Gamma/dm_{e^+e^-}$ is

\begin{equation} \label{eq11}
\begin{split}
\frac{d\Gamma}{dm_{e^+e^-}} = \frac{\alpha^2 C^2_{E0}}{12\pi\Lambda^4} ~~ m_{e^+e^-}  \left( \Delta m^2 - m^2_{e^+e^-}
\right)^\frac{3}{2}
\end{split}.
\end{equation}

The full width has an analytic solution in this case, 

\begin{equation} \label{FW_me}
\begin{split}
\Gamma_{E0}&=\frac{\alpha^2 C^2_{E0}}{12\pi\Lambda^4}~
\int_{2m_e}^{m-m_0}  m_{e^+e^-}  \left( \Delta m^2 - m^2_{e^+e^-}\right)^\frac{3}{2}dm_{e^+e^-}\\
                      &=\frac{\alpha^2 C^2_{E0}}{\Lambda^4}~
\frac {\left(\Delta m^2-4m_e^2\right)^\frac{5}{2}}{60\pi}
\approx \frac{\alpha^2 C^2_{E0}}{\Lambda^4} \frac {\Delta m^5}{60\pi}
\end{split}
\end{equation}
that with 2\% accuracy corresponds to the exact solution of Eq.~\ref{FW_2}.

The differential width $d\Gamma/dm_{e^+e^-}$  from Eq.~\ref{eq4}, integrated over $\cos{\theta^*}$ and normalized to one (shown as the blue curve on   Figure~\ref{fig:2_1}) is very smooth from the minimum value of $m_{e^+e^-}=2m_e$ up to the maximum of  $\Delta m=$20.21 MeV.  The ATOMKI signal cannot arise from the straightforward analysis carried out in this section.

\section{\texorpdfstring{Electromagnetic transition form factors $F(q^2)$ in the nuclear decay 
$\mathrm{^{4}He^*} \to \mathrm{^{4}He} + e^+e^-$}{Electromagnetic transition form factors F(q2) in the nuclear decay He* to He + e+e-}}

\begin{figure}[h]
\begin{center}
\includegraphics[width=0.5\textwidth]{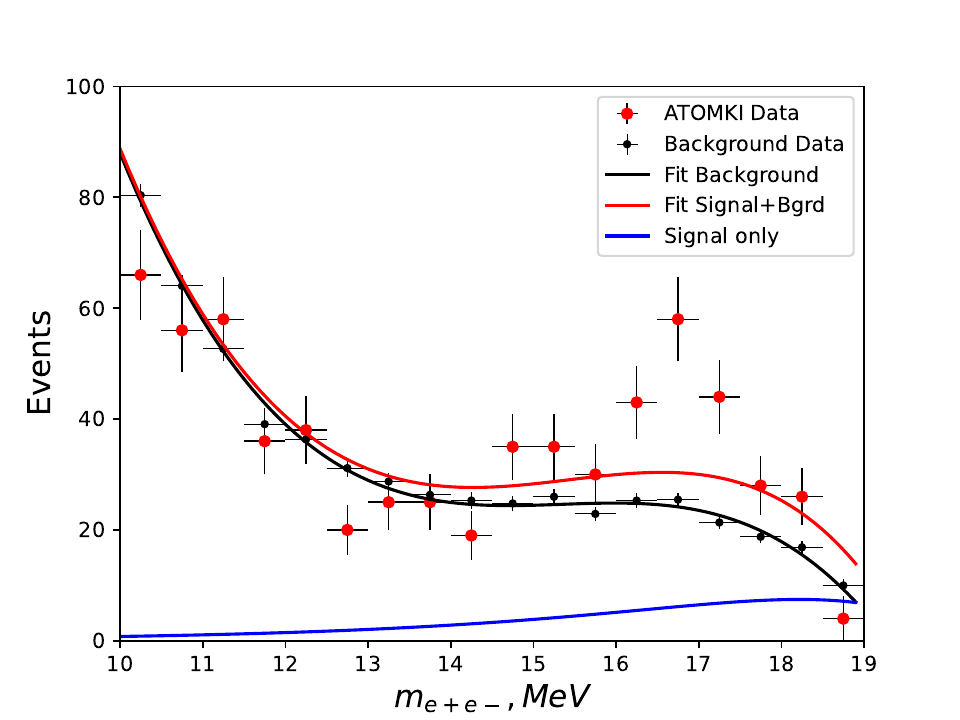}
\vspace*{-0.0cm}
\caption{
Fit to signal template (event counts vs. invariant mass distribution) for the transition ${\rm ^4He^*}(20.21)\to {\rm ^4He}+e^+e^-$.  The fit is an electromagnetic transition with form factor $F(q^2)\propto \exp{(q^2/\lambda^2)}$ (in red).  The background distribution is shown in black, the signal alone is shown in blue.
The fit is not satisfactory.}
\label{fig:2_2}
\end{center}
\end{figure}

\begin{figure}[h]
\begin{center}
\includegraphics[width=0.5\textwidth]{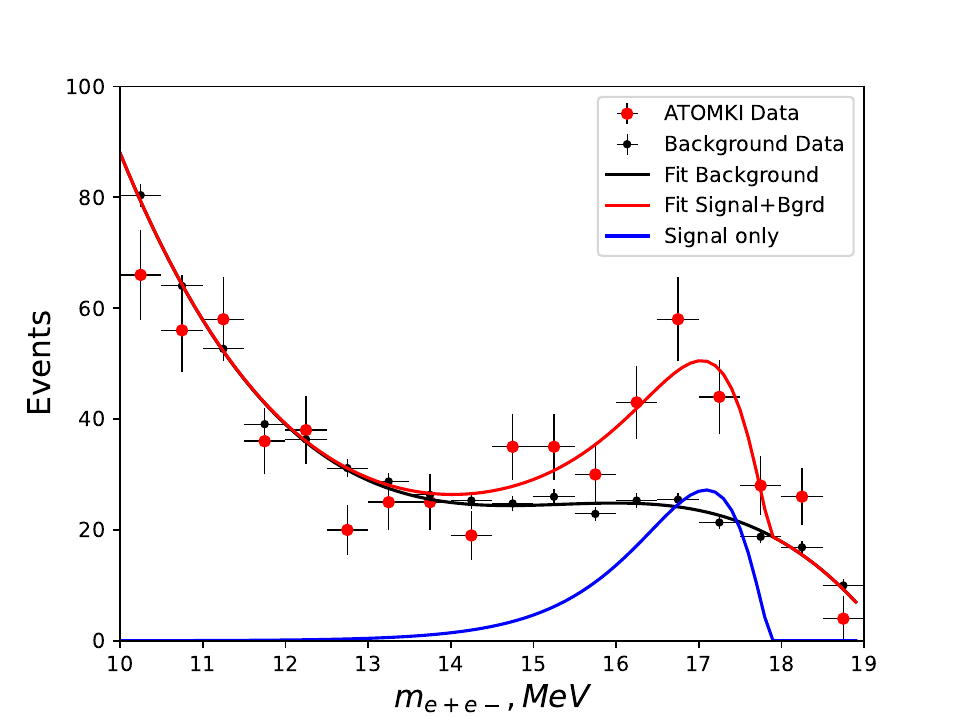}
\vspace*{-0.0cm}
\caption{Fit to signal template (event counts vs. invariant mass distribution) for the electromagnetic transition $\rm {HdQ^*}(17.9)\to{\rm ^4He}+(e^+e^-)$.  The fit assumes a form factor  $F(q^2)\propto \exp{(q^2/\lambda^2)}$ with the form factor parameter $\lambda$ = 6.2 MeV (in red).  The background distribution is shown in black, and the signal alone is shown in blue (see Eq.~\ref{eq12} with $\Delta m=17.9$ MeV).}
\label{fig:fig_x18}
\end{center}
\end{figure}

The analysis carried out in the previous section implicitly assumed a form factor $F(q^2)$ equal to unity.  A more realistic form factor significantly changes the $m_{e^+e^-}$ invariant mass distribution in the nuclear decay ${\rm ^4He}^*\to {\rm ^4He}+ e^+e^-$ as given by
\begin{equation} \label{eq12}
\begin{split}
\frac{d\Gamma}{dm_{e^+e^-}} = \frac{\alpha^2 C^2_{E0}}{12\pi\Lambda^4} ~~ m_{e^+e^-}  \left( \Delta m^2 - m^2_{e^+e^-}
\right)^\frac{3}{2} |F(q^2)|^2
\end{split}.
\end{equation}
\noindent We show the best fit form factor and its failure to describe the signal in this section, as well as the successful fit for a new excitation energy near $18$ MeV.
The experimental $m_{e^+e^-}$ distribution is not given in the ATOMKI work.  In order to compare the theoretical analysis in this section to the signal, we first extract the ATOMKI data from \cite{Krasznahorkay:2019lyl}, graph them and confirm that we can replicate their plot.  We then use this as a template to fit our signal, noting that it is based on the unpublished ATOMKI work \cite{Krasznahorkay:2019lyl} and cross-checked with the invariant mass plot in the NA64 Collaboration X17 simulations \cite{NA64:2020xxh}.  In the next step, we use the extracted signal and background distributions from the template and make a fit with two functions to describe both distributions.  
Figure~\ref{fig:fig_gauss0} presents our fit of ATOMKI data (shown in red). The background (shown in black) was fitted independently because the experiment measured the background distribution with much higher accuracy than the signal. We used a third-degree polynomial to fit the background. The resulting function is 
$f_{\rm bg}(x)=1584-306x+20x^2-0.434x^3$  (where $x=m_{e^+e^-}$) with $\chi^2/ndf=0.99$. 
The same function was used in all fits in Figure~\ref{fig:fig_gauss0}, Figure~\ref{fig:2_2} and Figure~\ref{fig:fig_x18}.
The signal in Figure~\ref{fig:fig_gauss0} was described by a Gaussian function with 3 parameters: the mass value of the peak position ($16.8\pm 0.23~\rm MeV$), the width ($\sigma=0.70\pm0.17~\rm MeV$) and overall normalization.  This fit is very close to the signal template.  We now model the results theoretically.

\begin{figure}[h]
\begin{center}
\includegraphics[width=0.5\textwidth]{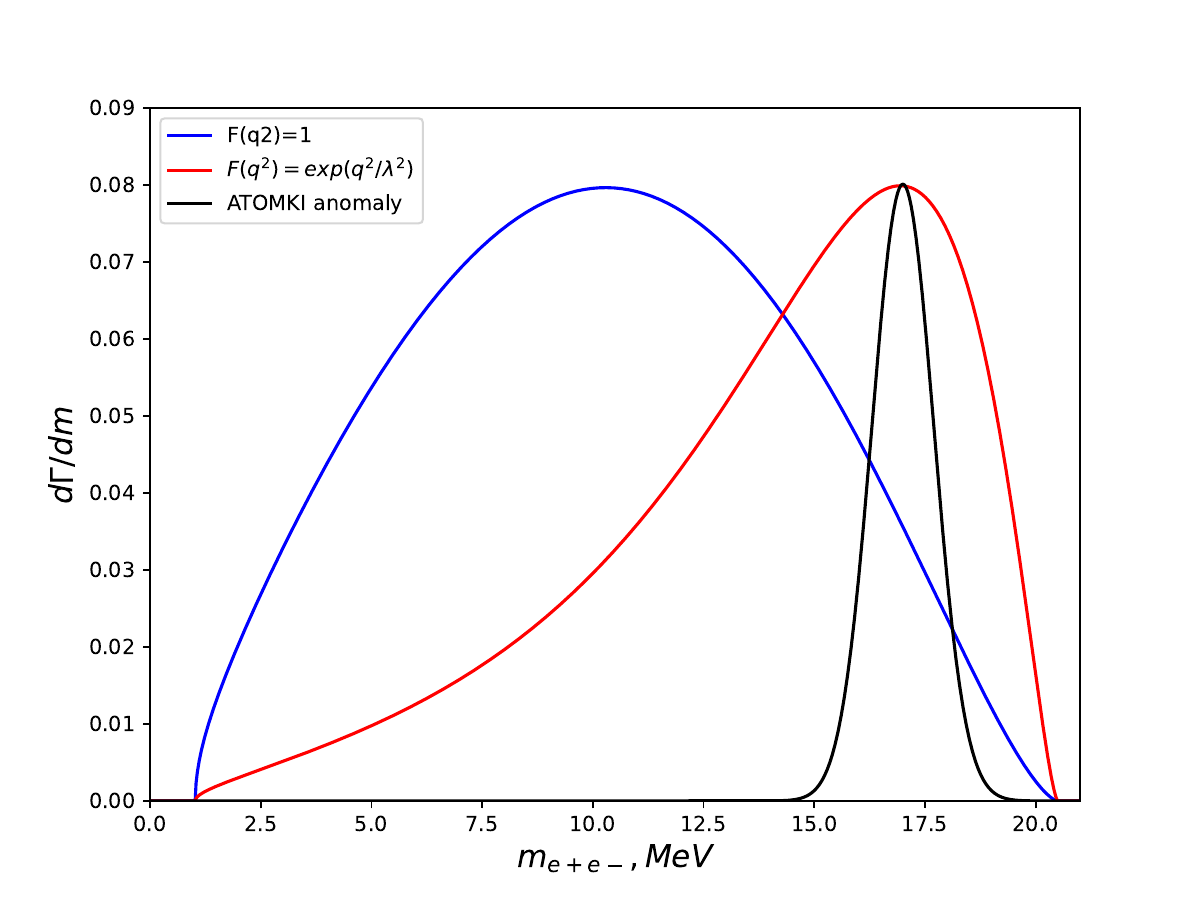}
\vspace*{-0.0cm}
\caption{Differential width $d\Gamma/dm_{e^+e^-}$ (in arbitrary units) for the transition ${\rm ^4He^*(20.21)}  \to   {\rm ^4He}+e^+e^-$ as a function of the electron-positron invariant mass $m_{e^+e^-}$. 
Blue curve has form factor $F(q^2)=1$ (see Eq.~\ref{eq12}).
The black curve is the Gaussian distribution of the ATOMKI anomaly with $m=$17 MeV and $\sigma=$0.7 MeV.
Red curve has form factor $F(q^2)\propto \exp{(q^2/\lambda^2)}$ with $\lambda=14.5$ MeV.
With this value of the $\lambda$ parameter, the distribution maximum is at $m=$17 MeV.
The functions are normalized such that the peak values are the same for all curves. Neither curve fits the signal. 
}
\label{fig:2_1}
\end{center}
\end{figure}

Several realistic form factor models are tested in our attempt to model the data template.  
We compare the shape 
of the differential width $d\Gamma/dm_{e^+e^-}$ for the electromagnetic transition with form factor $F(q^2)\propto \exp{(q^2/\lambda^2)}$ and $\lambda=14.5$ MeV (see Eq.~\ref{eq12}) to the ATOMKI anomaly with maximum near $17$ MeV as 
shown in  Figure~\ref{fig:2_1}.  We include a curve with form factor $F(q^2)=1$ for comparison.  We find the width of the distribution to be too large to obtain a satisfactory description of the experimental distribution.  The signal template is too narrow in comparison with the theoretical curve as shown in Figure~\ref{fig:2_1}, comparing the red curve (theoretical curve) and black curve (signal template). 
Figure \ref{fig:2_2} displays the attempt to fit the signal template, again with electromagnetic transition form factors $F(q^2)\propto \exp{(q^2/\lambda^2)}$. 
The background function is the same as in Figure~\ref{fig:fig_gauss0}. This fit has only two parameters,  $\lambda$ and the 
normalization of the theoretical function Eq.~\ref{eq12}.  The fit is not satisfactory.  A form factor of  $F(q^2)\propto q^2$ also fails to describe the signal due to its large width.

We now introduce a new parameter, the mass of a new excited $\rm ^4He^*$ state, into our model.  We motivate this state physically by the 12-quark color singlet Fock state model of $\rm ^4He$ nucleus in which the $\rm ^4He$ wavefunction is dominated by a linear combination of the nuclear state $|nnpp \rangle$ and the 12-quark QCD state $|[ud][ud][ud][ud][ud][ud] \rangle$ \cite{West:2020rlk}.
The result of the fit with form factor  $F(q^2)\propto \exp{(q^2/\lambda^2)}$ is presented in Figure~\ref{fig:fig_x18}. There are three parameters of the fit, the excitation energy of the proposed new $\rm {^4He^*}$ state $\rm E^*=17.9\pm 1$ MeV ($=\Delta m$ in Eq.~\ref{eq12}), the form factor parameter $\lambda=6.2\pm 0.5$ MeV  and the overall normalization. The model describes the data template well. The parameter $\lambda$ is close to the momentum scale found in earlier work on nuclear transitions \cite{Zhang:2017zap}.  

We have reproduced the ATOMKI signal with the introduction of a new excitation of $\rm^4He$ but there are serious physical issues that must be addressed.  First, the energy of the electron-positron pair in this case has to be close to $17.9~\rm MeV$. This is at the lower limit of the selection of the positrons and electrons with total energy in the region ($18,22$) MeV and $2.6 \sigma$ from the central excitation energy $20.49~\rm MeV$ created by the $900~\rm keV$ proton beam.  There are no error bars given by ATOMKI but a reasonable visual estimate is $0.9~\rm MeV$ which would reduce our $\sigma$ values. These are not prohibitively large deviations and we do not discuss them further but they must be noted.  More importantly, this is an excitation energy that has never been observed in a decay to a conventional hadronic or nuclear final state.  In addition, the form factor parameter $\lambda$ is an order of magnitude smaller than expected for a typical nuclear transition.  We address both of the latter points in the following section. 

\section{QCD Hexadiquark excitations}
\label{sec:hdq}
The hexadiquark (HdQ) is a hidden-color QCD state \cite{Brodsky:1976rz,Harvey:1980rva,Brodsky:1983vf,Harvey:1981udr,Bashkanov:2013cla,Miller:2013hla} in the $\rm ^4He$ nuclear wavefunction consisting of six scalar $[ud]$ diquarks \cite{Jaffe:2004ph,West:2020tyo} in a color singlet configuration \cite{West:2020rlk}.  Excitation of a subset of diquarks, either radial or angular momentum excitations, gives rise to nuclear transitions with unconventional decays.  In conventional nuclear physics, for example, the electron scattering experiment $e + {\rm{^4He}} \to e' + X$ will produce hadronic excitations of $\rm ^4He$. In contrast, if an electron (or proton) scatters on the hexadiquark Fock state of $\rm ^4He$, it can produce orbital or radial excitations between any two of the [ud] diquarks.  Typically such excitations would be at energies near $300~\rm MeV$ above the ground state, in analogy to the $N^*$ and $\Delta$ excited states of nucleons.  Excitations of single diquarks from spin-0 isospin-0 to spin-1 isospin-1 states have energies of approximately $200~\rm MeV$ \cite{West:2020tyo,Jaffe:2004ph}.

However, the HdQ is an unusual QCD eigenstate because, in addition to being a low mass spin-0 isospin-0 S-wave hidden-color state, the wavefunction contains three symmetric and  \textit{repulsive}  
$\bar 3_C \times \bar 3_C \rightarrow \bar 6_C$ interactions.  These interactions are required by the spin-statistics constraints upon clustering pairs of $[ud]$ diquarks into each of the three $[ud][ud]$ duo-diquarks contained within the 12-quark color-singlet wavefunction.  While the final construction of the three duo-diquarks into the HdQ is strongly bound with color factor $C_F=-5$,  the QCD interaction between diquarks in a duo-diquark is repulsive which significantly lowers the excitation energies of the HdQ and increases the size of the HdQ.

There are two ways to excite the $\rm 6_C$ bond between diquarks: an orbital angular momentum $\rm L = 1$ excitation and an $\rm L=0$ radial excitation.  These are color-singlet excitations, they do not break the strongly bound HdQ state apart, and therefore cannot decay to a conventional nuclear physics state such as $p + {\rm ^3H}$ or $n+ {\rm^3He}$.  However, since the $\rm L=1$ excitation has $J^{P}=1^{-}$, the HdQ excitation can decay to the $\rm^4He $ ground state by the emission of a virtual photon which makes an $e^+ e^-$  pair:  $\rm{^4He^*(17.9) \to {\rm ^4 He}}+ \gamma^* \to {\rm ^4He} + e^+e^-$.  This state can also emit a $17.9 \pm 1 ~\rm MeV$ real photon.  The $\rm L=0$ radial excitation between diquarks has $J^{P}=0^{+}$ and can only decay via virtual photon which creates the $e^+e^-$ pair.  Thus the three decay modes of the HdQ predict an $e^+e^-$ pair of invariant mass $17.9 \pm 1~\rm MeV$ and/or a $17.9 \pm 1~\rm MeV$ gamma ray photon.

In the case of an excitation of the $\rm ^4$He nucleus to a known excited state, there are two additional decay modes with distinct signatures.  In the first scenario, the transition is described by the  2-stage process
\begin{equation}
    {\rm p+^{3}H} \rightarrow{\rm ^4He^{*}}(20.21)~ \mathrm{MeV} \rightarrow {\rm ^4He^*}(17.9) ~\mathrm{MeV}+\gamma(2.31 \pm 1 ~ \mathrm{MeV}) \rightarrow{\rm ^4He}+\left(e^{+}e^{-}\right)+\gamma(2.31 \pm 1 ~ \mathrm{MeV})
\end{equation}
for the first excited state of $\rm ^4He$ at $20.21~\rm MeV$ ($\Gamma=0.50 ~\mathrm{MeV}$).  The second scenario occurs for the second excited state of $^4$He at $21.01$ MeV $(\Gamma =0.84 ~\mathrm{MeV})$,
\begin{equation}
    {\rm p+^{3}H} \rightarrow{\rm ^4He^{*}}(21.01)~ \mathrm{MeV} \rightarrow {\rm ^4He^*}(17.9) ~\mathrm{MeV}+\gamma(3.11 \pm 1 ~ \mathrm{MeV}) \rightarrow{\rm ^4He}+\left(e^{+}e^{-}\right)+\gamma(3.11 \pm 1 ~ \mathrm{MeV})
\end{equation}

\noindent It is important to note that multiple processes can excite the HdQ state, including direct proton capture in the initial $p+\rm ^3H$ collision creating $\gamma + {\rm HdQ^*}$ and then decaying down to $e^+e^- + {\rm ^4He}$, \textit{i.e.}, 

\begin{equation}
    p+{\rm ^3H} \rightarrow \gamma + {\rm ^4He^*}(17.9) \rightarrow {\rm ^4He} +  e^+e^-  + \gamma.
\end{equation}
A low energy photon will also be present in the final state, however, the signature of this decay is the $e^+e^-$ pair of invariant mass $17.9 \pm 1~\rm MeV$.  In addition to the decay signatures outlined above (summarized in Tables \ref{tab:sigs} and \ref{tab:sigs2}), a key method for confirmation of the hexadiquark solution to the ATOMKI puzzle is an investigation of the first excited state of $^4$He to identify a secondary ``shoulder'' peak within the range of $17.9 \pm 1~\rm MeV$.  Since the hexadiquark contains three sets of duo-diquarks, the $17.9 \pm 1~\rm MeV$ excitation will be accompanied by excitations at $\sim 36$ MeV and $\sim 54$ MeV.  Larger $\alpha$-nuclei, including $^8$Be and $^{12}$C, will have even higher multiples of the $17.9 \pm 1~\rm MeV$ excitation for each additional hexadiquark state within the nuclear wavefunction.

We propose that the HdQ is the origin of a previously missed subdominant $\rm ^4He^*(17.9)$ excited state.  It can be observed in the interaction $e + {\rm{^4He}}  \to e' + {\rm{^4He^*}}(17.9)  \to e' + {\rm{^4He}} + [e^+e^-]$.  The characteristic mass scale of the transition form factor $F(q^2)$ producing the timelike lepton pair in the $\rm HdQ^* \to HdQ + \gamma^*$ process can be significantly lighter than the standard QCD mass scale because it reflects the large size structure of the HdQ due to the repulsive interactions within each of the three duo-diquarks.  This can cause a $\lambda$ parameter of order $10$ MeV in the transition $\rm HdQ^* \to HdQ$ form factor as required by our fit.  The conclusion from these arguments is that an excitation of the HdQ Fock state of $\rm ^4He$ can account for the excellent fit and parameters in Figure~\ref{fig:fig_x18} to the ATOMKI signal template. 
  
We note that the hexadiquark state is predicted to exist as a hidden-color QCD Fock state within every nucleus containing an alpha particle. Each such nucleus will also have HdQ hidden color excitations; the resulting $\rm HdQ^*$ decays should therefore also be observed in lepton pair decays from excitations of $\rm ^8Be$ and $\rm ^{12}C$.  All nuclei containing at least one $\alpha$-particle will have this excitation and the strength of the signal is predicted to increase with each additional $\alpha$-particle.  HdQ transitions cannot occur in $A=3$ nuclei, in deuterium, or in nucleons (all of which contain less than 12 quarks) and therefore will not be observed in experiments that use these targets.  However, other hidden-color Fock states in $A< 4$ nuclei that use the repulsive $6_C$ representation of $\rm SU(3)_C$ also allow for new lower energy excitations with unconventional decays.

There are important complementary tests for the hexadiquark solution. The HdQ component of $\rm ^4He$ should be observed as a peak in the missing mass in $e+{\rm{^4He}} \to e'+ X$.  The diquark composition of the ${\rm{^4He}}$ nucleus may be observed via $e+ {\rm{^4He}} \to e'+[ud]+X$  where the $[ud]$-diquark jet \cite{Ilgenfritz:1978cc} is produced opposite to the scattered electron.

\begin{table}[htbp]
\centering
\begin{tabular}{|c|c|c|}
\hline
\textbf{New Hexadiquark excitation}  & \textbf{Decay channel } & \textbf{Experimental signature}\\
\textbf{$^4\rm He^*(17.9)$}  & \textbf{$^4\rm He^*(17.9 )\to $} & \textbf{}\\
\hline
L=0 radial excitations $[ud]\leftrightarrow [ud]$ & $^4$He $+ \gamma^*$ $\rightarrow$ $^4$He + $e^+e^-$ & $e^+e^-(17.9)$\\
\hline 
L=1 orbital excitations $[ud] \circlearrowright [ud]$ &  $^4$He + $\gamma$ & $\gamma(17.9)$\\
~ & $^4$He + $\gamma \rightarrow$ $^4$He + $ e^+e^-$ & $e^+e^-(17.9)$\\
\hline
\end{tabular}
\caption{Experimental signatures for direct excitations of the QCD hexadiquark Fock state in units of MeV with uncertainties of $\sim 1$ MeV.  Each hexadiquark contains 3 diquark-diquark correlations and therefore subdominant nuclear excitations near 36 MeV and 54 MeV are predicted in $^4$He.  $^8$Be, $^{12}$C and all higher A $\alpha$-cluster nuclei will have additional excitations at multiples of $17.9 $ MeV, exactly 3 additional excitations per $\alpha$.
\label{tab:sigs}}
\end{table}

\begin{table}[htbp]
\centering
\begin{tabular}{|c|c|c|}
\hline
\textbf{Observed Excitation} & \textbf{Decay channel $^4\rm He^*(E^*)\to $} & \textbf{Experimental}\\
\textbf{energy of $^4\rm He$} & \textbf{} & \textbf{signature}\\
\hline
$^4\rm He^*(20.21)$ & ${\rm ^4He}^*(17.9) $+ $\gamma(2.31)$ $\rightarrow {\rm ^4He} + \gamma(2.31) + e^+e^-(17.9)$ & $e^+e^-(17.9)$, $\gamma(2.31)$\\
~ &  ${\rm ^4He}^{*}(17.9)+\gamma(2.31)\rightarrow {\rm ^4He} + \gamma(17.9)$ & $\gamma(17.9)$, $\gamma(2.31 )$ \\
\hline 
$^4\rm He^*(21.01)$ & ${\rm ^4He}^*(17.9)$+ $\gamma(3.11)$ $\rightarrow {\rm ^4He} + \gamma(3.11)+ e^+e^-(17.9)$ & $e^+e^-(17.9)$, $\gamma(3.11)$\\
~ & ${\rm ^4He}^*(17.9)$+ $\gamma(3.11)$ $\rightarrow {\rm ^4He} + \gamma(3.11)+ \gamma(17.9)$ & $\gamma(17.9)$, $\gamma(3.11)$\\
\hline
\end{tabular}
\caption{Experimental signatures for 2-stage excitations in which excitation of a known ${\rm ^4He}$ excited state is followed by a decay to the proposed new hexadiquark excitation, $^4 \rm He^*\rightarrow {\rm He^*(17.9)}\rightarrow X$, in units of MeV. Uncertainties are $\sim 1$ MeV. \label{tab:sigs2}}
\end{table}

\section{Summary}
We have presented a QCD-based solution to the ATOMKI anomalies that utilizes the predicted hexadiquark hidden-color Fock state in the $\rm ^4He$  nuclear wavefunction \cite{West:2020rlk}.  The hexadiquark state mixes with the conventional $|nnpp\rangle$ nuclear physics state of $\rm ^4He$ and offers a hidden-color explanation of its anomalously large binding energy.  Our calculation of the $m_{e^+e^-}$ invariant mass spectrum in the electromagnetic transition ${\rm ^4He}^*({\rm E^*})\to {\rm ^4He}+ e^+e^-$ as well as the observed differential and total width of the decay are fit to an electromagnetic decay of a new excitation of $\rm ^4He$ of energy $\rm E^*=17.9 \pm 1 ~\rm MeV$ above the ground state.  The source of $\rm ^4He^*(17.9)$ is proposed to be low energy diquark-diquark pair excitations within the hidden-color hexadiquark component of the nuclear wavefunction, enabled by the novel spin-statistics mandated use of weakly repulsive $\rm 6_C$ interactions of $\rm SU(3)_C$ between diquark pairs.

We have shown that it is possible to reproduce the ATOMKI signal with ${\rm J^{P}}=0^{+}$ and ${\rm J^{P}}=1^{-}$ excitations between the diquark constituents of the hexadiquark Fock state.  In one production scenario, the proton $+$ tritium collision first produces the conventional first nuclear excited state $\rm ^4He^*$ which decays by photon emission to an excitation of the QCD Fock state.  The hexadiquark excitation may be produced by direct proton capture as well.  All experimental signatures and decay channels are summarized in Tables  \ref{tab:sigs} and \ref{tab:sigs2}.  

This work is both timely and important due to the negative searches for the new X17 boson carried out thus far \cite{MEGII:2024urz,NA482:2015wmo,NA64:2019auh,NA64:2020xxh} and planned searches at Jefferson Lab \cite{Ahmidouch:2021edo}, in Italy with the Positron Annihilation to Dark Matter Experiment (PADME \cite{PADME:2022tqr,Darme:2022zfw}) and elsewhere.  Our QCD-based solution of the ATOMKI results, which must occur in all $A\geq4$ nuclei including the $^4\rm He$, $^8 \rm Be$, $^{12}\rm C$ nuclei in the ATOMKI experiments, is in contrast to the new particle explanation.

\section*{Acknowledgements}
J.R.W. acknowledges support by the LDRD program of LBNL, the EIC Center at Jefferson Lab and the U.S. Department of Energy, Office of Science, Office of Nuclear Physics, under contract number DE-AC02-05CH11231.  V.K. acknowledges support by the U.S. Department of Energy, Office of Science, and Office of Nuclear Physics under contracts DE-AC05-06OR23177. JLAB-PHY-22-3641.  The work of S.J.B. was supported in part by the Department of Energy under Contract No. DE-AC02- 76SF00515. SLAC-PUB-17683.

\bibliography{atomki}

\begin{thebibliography}{55}%
\makeatletter
\providecommand \@ifxundefined [1]{%
 \@ifx{#1\undefined}
}%
\providecommand \@ifnum [1]{%
 \ifnum #1\expandafter \@firstoftwo
 \else \expandafter \@secondoftwo
 \fi
}%
\providecommand \@ifx [1]{%
 \ifx #1\expandafter \@firstoftwo
 \else \expandafter \@secondoftwo
 \fi
}%
\providecommand \natexlab [1]{#1}%
\providecommand \enquote  [1]{``#1''}%
\providecommand \bibnamefont  [1]{#1}%
\providecommand \bibfnamefont [1]{#1}%
\providecommand \citenamefont [1]{#1}%
\providecommand \href@noop [0]{\@secondoftwo}%
\providecommand \href [0]{\begingroup \@sanitize@url \@href}%
\providecommand \@href[1]{\@@startlink{#1}\@@href}%
\providecommand \@@href[1]{\endgroup#1\@@endlink}%
\providecommand \@sanitize@url [0]{\catcode `\\12\catcode `\$12\catcode
  `\&12\catcode `\#12\catcode `\^12\catcode `\_12\catcode `\%12\relax}%
\providecommand \@@startlink[1]{}%
\providecommand \@@endlink[0]{}%
\providecommand \url  [0]{\begingroup\@sanitize@url \@url }%
\providecommand \@url [1]{\endgroup\@href {#1}{\urlprefix }}%
\providecommand \urlprefix  [0]{URL }%
\providecommand \Eprint [0]{\href }%
\providecommand \doibase [0]{http://dx.doi.org/}%
\providecommand \selectlanguage [0]{\@gobble}%
\providecommand \bibinfo  [0]{\@secondoftwo}%
\providecommand \bibfield  [0]{\@secondoftwo}%
\providecommand \translation [1]{[#1]}%
\providecommand \BibitemOpen [0]{}%
\providecommand \bibitemStop [0]{}%
\providecommand \bibitemNoStop [0]{.\EOS\space}%
\providecommand \EOS [0]{\spacefactor3000\relax}%
\providecommand \BibitemShut  [1]{\csname bibitem#1\endcsname}%
\let\auto@bib@innerbib\@empty
\bibitem [{\citenamefont {Krasznahorkay}\ \emph {et~al.}(2021)\citenamefont
  {Krasznahorkay}, \citenamefont {Csatl\'os}, \citenamefont {Csige},
  \citenamefont {Guly\'as}, \citenamefont {Krasznahorkay}, \citenamefont
  {Nyak\'o}, \citenamefont {Rajta}, \citenamefont {Tim\'ar}, \citenamefont
  {Vajda},\ and\ \citenamefont {Sas}}]{Krasznahorkay:2021joi}%
  \BibitemOpen
  \bibfield  {author} {\bibinfo {author} {\bibfnamefont {A.~J.}\ \bibnamefont
  {Krasznahorkay}}, \bibinfo {author} {\bibfnamefont {M.}~\bibnamefont
  {Csatl\'os}}, \bibinfo {author} {\bibfnamefont {L.}~\bibnamefont {Csige}},
  \bibinfo {author} {\bibfnamefont {J.}~\bibnamefont {Guly\'as}}, \bibinfo
  {author} {\bibfnamefont {A.}~\bibnamefont {Krasznahorkay}}, \bibinfo {author}
  {\bibfnamefont {B.~M.}\ \bibnamefont {Nyak\'o}}, \bibinfo {author}
  {\bibfnamefont {I.}~\bibnamefont {Rajta}}, \bibinfo {author} {\bibfnamefont
  {J.}~\bibnamefont {Tim\'ar}}, \bibinfo {author} {\bibfnamefont
  {I.}~\bibnamefont {Vajda}}, \ and\ \bibinfo {author} {\bibfnamefont {N.~J.}\
  \bibnamefont {Sas}},\ }\href {\doibase 10.1103/PhysRevC.104.044003}
  {\bibfield  {journal} {\bibinfo  {journal} {Phys. Rev. C}\ }\textbf {\bibinfo
  {volume} {104}},\ \bibinfo {pages} {044003} (\bibinfo {year} {2021})},\
  \Eprint {http://arxiv.org/abs/2104.10075} {arXiv:2104.10075 [nucl-ex]}
  \BibitemShut {NoStop}%
\bibitem [{\citenamefont {Krasznahorkay}\ \emph {et~al.}(2019)\citenamefont
  {Krasznahorkay} \emph {et~al.}}]{Krasznahorkay:2019lyl}%
  \BibitemOpen
  \bibfield  {author} {\bibinfo {author} {\bibfnamefont {A.~J.}\ \bibnamefont
  {Krasznahorkay}} \emph {et~al.},\ }\href@noop {} {\  (\bibinfo {year}
  {2019})},\ \Eprint {http://arxiv.org/abs/1910.10459} {arXiv:1910.10459
  [nucl-ex]} \BibitemShut {NoStop}%
\bibitem [{\citenamefont {Krasznahorkay}\ \emph {et~al.}(2016)\citenamefont
  {Krasznahorkay} \emph {et~al.}}]{Krasznahorkay:2015iga}%
  \BibitemOpen
  \bibfield  {author} {\bibinfo {author} {\bibfnamefont {A.~J.}\ \bibnamefont
  {Krasznahorkay}} \emph {et~al.},\ }\href {\doibase
  10.1103/PhysRevLett.116.042501} {\bibfield  {journal} {\bibinfo  {journal}
  {Phys. Rev. Lett.}\ }\textbf {\bibinfo {volume} {116}},\ \bibinfo {pages}
  {042501} (\bibinfo {year} {2016})},\ \Eprint
  {http://arxiv.org/abs/1504.01527} {arXiv:1504.01527 [nucl-ex]} \BibitemShut
  {NoStop}%
\bibitem [{\citenamefont {Krasznahorkay}\ \emph {et~al.}(2022)\citenamefont
  {Krasznahorkay} \emph {et~al.}}]{Krasznahorkay:2022pxs}%
  \BibitemOpen
  \bibfield  {author} {\bibinfo {author} {\bibfnamefont {A.~J.}\ \bibnamefont
  {Krasznahorkay}} \emph {et~al.},\ }\href {\doibase
  10.1103/PhysRevC.106.L061601} {\bibfield  {journal} {\bibinfo  {journal}
  {Phys. Rev. C}\ }\textbf {\bibinfo {volume} {106}},\ \bibinfo {pages}
  {L061601} (\bibinfo {year} {2022})},\ \Eprint
  {http://arxiv.org/abs/2209.10795} {arXiv:2209.10795 [nucl-ex]} \BibitemShut
  {NoStop}%
\bibitem [{\citenamefont {Sas}\ \emph {et~al.}(2022)\citenamefont {Sas} \emph
  {et~al.}}]{Sas:2022pgm}%
  \BibitemOpen
  \bibfield  {author} {\bibinfo {author} {\bibfnamefont {N.~J.}\ \bibnamefont
  {Sas}} \emph {et~al.},\ }\href@noop {} {\  (\bibinfo {year} {2022})},\
  \Eprint {http://arxiv.org/abs/2205.07744} {arXiv:2205.07744 [nucl-ex]}
  \BibitemShut {NoStop}%
\bibitem [{\citenamefont {Battaglieri}\ \emph {et~al.}(2017)\citenamefont
  {Battaglieri} \emph {et~al.}}]{Battaglieri:2017aum}%
  \BibitemOpen
  \bibfield  {author} {\bibinfo {author} {\bibfnamefont {M.}~\bibnamefont
  {Battaglieri}} \emph {et~al.},\ }in\ \href@noop {} {\emph {\bibinfo
  {booktitle} {{U.S. Cosmic Visions: New Ideas in Dark Matter}}}}\ (\bibinfo
  {year} {2017})\ \Eprint {http://arxiv.org/abs/1707.04591} {arXiv:1707.04591
  [hep-ph]} \BibitemShut {NoStop}%
\bibitem [{\citenamefont {Backens}\ and\ \citenamefont
  {Vanderhaeghen}(2022)}]{Backens:2021qkv}%
  \BibitemOpen
  \bibfield  {author} {\bibinfo {author} {\bibfnamefont {J.}~\bibnamefont
  {Backens}}\ and\ \bibinfo {author} {\bibfnamefont {M.}~\bibnamefont
  {Vanderhaeghen}},\ }\href {\doibase 10.1103/PhysRevLett.128.091802}
  {\bibfield  {journal} {\bibinfo  {journal} {Phys. Rev. Lett.}\ }\textbf
  {\bibinfo {volume} {128}},\ \bibinfo {pages} {091802} (\bibinfo {year}
  {2022})},\ \Eprint {http://arxiv.org/abs/2110.06055} {arXiv:2110.06055
  [hep-ph]} \BibitemShut {NoStop}%
\bibitem [{\citenamefont {Fornal}(2017)}]{Fornal:2017msy}%
  \BibitemOpen
  \bibfield  {author} {\bibinfo {author} {\bibfnamefont {B.}~\bibnamefont
  {Fornal}},\ }\href {\doibase 10.1142/S0217751X17300204} {\bibfield  {journal}
  {\bibinfo  {journal} {Int. J. Mod. Phys. A}\ }\textbf {\bibinfo {volume}
  {32}},\ \bibinfo {pages} {1730020} (\bibinfo {year} {2017})},\ \Eprint
  {http://arxiv.org/abs/1707.09749} {arXiv:1707.09749 [hep-ph]} \BibitemShut
  {NoStop}%
\bibitem [{\citenamefont {Feng}\ \emph {et~al.}(2017)\citenamefont {Feng},
  \citenamefont {Fornal}, \citenamefont {Galon}, \citenamefont {Gardner},
  \citenamefont {Smolinsky}, \citenamefont {Tait},\ and\ \citenamefont
  {Tanedo}}]{Feng:2016ysn}%
  \BibitemOpen
  \bibfield  {author} {\bibinfo {author} {\bibfnamefont {J.~L.}\ \bibnamefont
  {Feng}}, \bibinfo {author} {\bibfnamefont {B.}~\bibnamefont {Fornal}},
  \bibinfo {author} {\bibfnamefont {I.}~\bibnamefont {Galon}}, \bibinfo
  {author} {\bibfnamefont {S.}~\bibnamefont {Gardner}}, \bibinfo {author}
  {\bibfnamefont {J.}~\bibnamefont {Smolinsky}}, \bibinfo {author}
  {\bibfnamefont {T.~M.~P.}\ \bibnamefont {Tait}}, \ and\ \bibinfo {author}
  {\bibfnamefont {P.}~\bibnamefont {Tanedo}},\ }\href {\doibase
  10.1103/PhysRevD.95.035017} {\bibfield  {journal} {\bibinfo  {journal} {Phys.
  Rev. D}\ }\textbf {\bibinfo {volume} {95}},\ \bibinfo {pages} {035017}
  (\bibinfo {year} {2017})},\ \Eprint {http://arxiv.org/abs/1608.03591}
  {arXiv:1608.03591 [hep-ph]} \BibitemShut {NoStop}%
\bibitem [{\citenamefont {Viviani}\ \emph {et~al.}(2022)\citenamefont
  {Viviani}, \citenamefont {Filandri}, \citenamefont {Girlanda}, \citenamefont
  {Gustavino}, \citenamefont {Kievsky}, \citenamefont {Marcucci},\ and\
  \citenamefont {Schiavilla}}]{Viviani:2021stx}%
  \BibitemOpen
  \bibfield  {author} {\bibinfo {author} {\bibfnamefont {M.}~\bibnamefont
  {Viviani}}, \bibinfo {author} {\bibfnamefont {E.}~\bibnamefont {Filandri}},
  \bibinfo {author} {\bibfnamefont {L.}~\bibnamefont {Girlanda}}, \bibinfo
  {author} {\bibfnamefont {C.}~\bibnamefont {Gustavino}}, \bibinfo {author}
  {\bibfnamefont {A.}~\bibnamefont {Kievsky}}, \bibinfo {author} {\bibfnamefont
  {L.~E.}\ \bibnamefont {Marcucci}}, \ and\ \bibinfo {author} {\bibfnamefont
  {R.}~\bibnamefont {Schiavilla}},\ }\href {\doibase
  10.1103/PhysRevC.105.014001} {\bibfield  {journal} {\bibinfo  {journal}
  {Phys. Rev. C}\ }\textbf {\bibinfo {volume} {105}},\ \bibinfo {pages}
  {014001} (\bibinfo {year} {2022})},\ \Eprint
  {http://arxiv.org/abs/2104.07808} {arXiv:2104.07808 [nucl-th]} \BibitemShut
  {NoStop}%
\bibitem [{\citenamefont {Agrawal}\ \emph {et~al.}(2021)\citenamefont {Agrawal}
  \emph {et~al.}}]{Agrawal:2021dbo}%
  \BibitemOpen
  \bibfield  {author} {\bibinfo {author} {\bibfnamefont {P.}~\bibnamefont
  {Agrawal}} \emph {et~al.},\ }\href {\doibase 10.1140/epjc/s10052-021-09703-7}
  {\bibfield  {journal} {\bibinfo  {journal} {Eur. Phys. J. C}\ }\textbf
  {\bibinfo {volume} {81}},\ \bibinfo {pages} {1015} (\bibinfo {year}
  {2021})},\ \Eprint {http://arxiv.org/abs/2102.12143} {arXiv:2102.12143
  [hep-ph]} \BibitemShut {NoStop}%
\bibitem [{\citenamefont {Alexander}\ \emph {et~al.}(2016)\citenamefont
  {Alexander} \emph {et~al.}}]{Alexander:2016aln}%
  \BibitemOpen
  \bibfield  {author} {\bibinfo {author} {\bibfnamefont {J.}~\bibnamefont
  {Alexander}} \emph {et~al.}\ }(\bibinfo {year} {2016})\ \Eprint
  {http://arxiv.org/abs/1608.08632} {arXiv:1608.08632 [hep-ph]} \BibitemShut
  {NoStop}%
\bibitem [{\citenamefont {Kozaczuk}\ \emph {et~al.}(2017)\citenamefont
  {Kozaczuk}, \citenamefont {Morrissey},\ and\ \citenamefont
  {Stroberg}}]{Kozaczuk:2016nma}%
  \BibitemOpen
  \bibfield  {author} {\bibinfo {author} {\bibfnamefont {J.}~\bibnamefont
  {Kozaczuk}}, \bibinfo {author} {\bibfnamefont {D.~E.}\ \bibnamefont
  {Morrissey}}, \ and\ \bibinfo {author} {\bibfnamefont {S.~R.}\ \bibnamefont
  {Stroberg}},\ }\href {\doibase 10.1103/PhysRevD.95.115024} {\bibfield
  {journal} {\bibinfo  {journal} {Phys. Rev. D}\ }\textbf {\bibinfo {volume}
  {95}},\ \bibinfo {pages} {115024} (\bibinfo {year} {2017})},\ \Eprint
  {http://arxiv.org/abs/1612.01525} {arXiv:1612.01525 [hep-ph]} \BibitemShut
  {NoStop}%
\bibitem [{\citenamefont {Kitahara}\ and\ \citenamefont
  {Yamamoto}(2017)}]{Kitahara:2016zyb}%
  \BibitemOpen
  \bibfield  {author} {\bibinfo {author} {\bibfnamefont {T.}~\bibnamefont
  {Kitahara}}\ and\ \bibinfo {author} {\bibfnamefont {Y.}~\bibnamefont
  {Yamamoto}},\ }\href {\doibase 10.1103/PhysRevD.95.015008} {\bibfield
  {journal} {\bibinfo  {journal} {Phys. Rev. D}\ }\textbf {\bibinfo {volume}
  {95}},\ \bibinfo {pages} {015008} (\bibinfo {year} {2017})},\ \Eprint
  {http://arxiv.org/abs/1609.01605} {arXiv:1609.01605 [hep-ph]} \BibitemShut
  {NoStop}%
\bibitem [{\citenamefont {Feng}\ \emph {et~al.}(2020)\citenamefont {Feng},
  \citenamefont {Tait},\ and\ \citenamefont {Verhaaren}}]{Feng:2020mbt}%
  \BibitemOpen
  \bibfield  {author} {\bibinfo {author} {\bibfnamefont {J.~L.}\ \bibnamefont
  {Feng}}, \bibinfo {author} {\bibfnamefont {T.~M.~P.}\ \bibnamefont {Tait}}, \
  and\ \bibinfo {author} {\bibfnamefont {C.~B.}\ \bibnamefont {Verhaaren}},\
  }\href {\doibase 10.1103/PhysRevD.102.036016} {\bibfield  {journal} {\bibinfo
   {journal} {Phys. Rev. D}\ }\textbf {\bibinfo {volume} {102}},\ \bibinfo
  {pages} {036016} (\bibinfo {year} {2020})},\ \Eprint
  {http://arxiv.org/abs/2006.01151} {arXiv:2006.01151 [hep-ph]} \BibitemShut
  {NoStop}%
\bibitem [{\citenamefont {Zhang}\ and\ \citenamefont
  {Miller}(2017)}]{Zhang:2017zap}%
  \BibitemOpen
  \bibfield  {author} {\bibinfo {author} {\bibfnamefont {X.}~\bibnamefont
  {Zhang}}\ and\ \bibinfo {author} {\bibfnamefont {G.~A.}\ \bibnamefont
  {Miller}},\ }\href {\doibase 10.1016/j.physletb.2017.08.013} {\bibfield
  {journal} {\bibinfo  {journal} {Phys. Lett. B}\ }\textbf {\bibinfo {volume}
  {773}},\ \bibinfo {pages} {159} (\bibinfo {year} {2017})},\ \Eprint
  {http://arxiv.org/abs/1703.04588} {arXiv:1703.04588 [nucl-th]} \BibitemShut
  {NoStop}%
\bibitem [{\citenamefont {Zhang}\ and\ \citenamefont
  {Miller}(2021)}]{Zhang:2020ukq}%
  \BibitemOpen
  \bibfield  {author} {\bibinfo {author} {\bibfnamefont {X.}~\bibnamefont
  {Zhang}}\ and\ \bibinfo {author} {\bibfnamefont {G.~A.}\ \bibnamefont
  {Miller}},\ }\href {\doibase 10.1016/j.physletb.2021.136061} {\bibfield
  {journal} {\bibinfo  {journal} {Phys. Lett. B}\ }\textbf {\bibinfo {volume}
  {813}},\ \bibinfo {pages} {136061} (\bibinfo {year} {2021})},\ \Eprint
  {http://arxiv.org/abs/2008.11288} {arXiv:2008.11288 [hep-ph]} \BibitemShut
  {NoStop}%
\bibitem [{\citenamefont {Frugiuele}\ \emph {et~al.}(2017)\citenamefont
  {Frugiuele}, \citenamefont {Fuchs}, \citenamefont {Perez},\ and\
  \citenamefont {Schlaffer}}]{Frugiuele:2016rii}%
  \BibitemOpen
  \bibfield  {author} {\bibinfo {author} {\bibfnamefont {C.}~\bibnamefont
  {Frugiuele}}, \bibinfo {author} {\bibfnamefont {E.}~\bibnamefont {Fuchs}},
  \bibinfo {author} {\bibfnamefont {G.}~\bibnamefont {Perez}}, \ and\ \bibinfo
  {author} {\bibfnamefont {M.}~\bibnamefont {Schlaffer}},\ }\href {\doibase
  10.1103/PhysRevD.96.015011} {\bibfield  {journal} {\bibinfo  {journal} {Phys.
  Rev. D}\ }\textbf {\bibinfo {volume} {96}},\ \bibinfo {pages} {015011}
  (\bibinfo {year} {2017})},\ \Eprint {http://arxiv.org/abs/1602.04822}
  {arXiv:1602.04822 [hep-ph]} \BibitemShut {NoStop}%
\bibitem [{\citenamefont {Banerjee}\ \emph {et~al.}(2018)\citenamefont
  {Banerjee} \emph {et~al.}}]{NA64:2018lsq}%
  \BibitemOpen
  \bibfield  {author} {\bibinfo {author} {\bibfnamefont {D.}~\bibnamefont
  {Banerjee}} \emph {et~al.} (\bibinfo {collaboration} {NA64}),\ }\href
  {\doibase 10.1103/PhysRevLett.120.231802} {\bibfield  {journal} {\bibinfo
  {journal} {Phys. Rev. Lett.}\ }\textbf {\bibinfo {volume} {120}},\ \bibinfo
  {pages} {231802} (\bibinfo {year} {2018})},\ \Eprint
  {http://arxiv.org/abs/1803.07748} {arXiv:1803.07748 [hep-ex]} \BibitemShut
  {NoStop}%
\bibitem [{\citenamefont {Castro}\ and\ \citenamefont
  {Quintero}(2021)}]{Castro:2021gdf}%
  \BibitemOpen
  \bibfield  {author} {\bibinfo {author} {\bibfnamefont {G.~L.}\ \bibnamefont
  {Castro}}\ and\ \bibinfo {author} {\bibfnamefont {N.}~\bibnamefont
  {Quintero}},\ }\href {\doibase 10.1103/PhysRevD.103.093002} {\bibfield
  {journal} {\bibinfo  {journal} {Phys. Rev. D}\ }\textbf {\bibinfo {volume}
  {103}},\ \bibinfo {pages} {093002} (\bibinfo {year} {2021})},\ \Eprint
  {http://arxiv.org/abs/2101.01865} {arXiv:2101.01865 [hep-ph]} \BibitemShut
  {NoStop}%
\bibitem [{\citenamefont {Dror}\ \emph {et~al.}(2017)\citenamefont {Dror},
  \citenamefont {Lasenby},\ and\ \citenamefont {Pospelov}}]{Dror:2017ehi}%
  \BibitemOpen
  \bibfield  {author} {\bibinfo {author} {\bibfnamefont {J.~A.}\ \bibnamefont
  {Dror}}, \bibinfo {author} {\bibfnamefont {R.}~\bibnamefont {Lasenby}}, \
  and\ \bibinfo {author} {\bibfnamefont {M.}~\bibnamefont {Pospelov}},\ }\href
  {\doibase 10.1103/PhysRevLett.119.141803} {\bibfield  {journal} {\bibinfo
  {journal} {Phys. Rev. Lett.}\ }\textbf {\bibinfo {volume} {119}},\ \bibinfo
  {pages} {141803} (\bibinfo {year} {2017})},\ \Eprint
  {http://arxiv.org/abs/1705.06726} {arXiv:1705.06726 [hep-ph]} \BibitemShut
  {NoStop}%
\bibitem [{\citenamefont {Alves}\ and\ \citenamefont
  {Weiner}(2018)}]{Alves:2017avw}%
  \BibitemOpen
  \bibfield  {author} {\bibinfo {author} {\bibfnamefont {D.~S.~M.}\
  \bibnamefont {Alves}}\ and\ \bibinfo {author} {\bibfnamefont
  {N.}~\bibnamefont {Weiner}},\ }\href {\doibase 10.1007/JHEP07(2018)092}
  {\bibfield  {journal} {\bibinfo  {journal} {JHEP}\ }\textbf {\bibinfo
  {volume} {07}},\ \bibinfo {pages} {092} (\bibinfo {year} {2018})},\ \Eprint
  {http://arxiv.org/abs/1710.03764} {arXiv:1710.03764 [hep-ph]} \BibitemShut
  {NoStop}%
\bibitem [{\citenamefont {Ariga}\ \emph {et~al.}(2019)\citenamefont {Ariga}
  \emph {et~al.}}]{FASER:2018eoc}%
  \BibitemOpen
  \bibfield  {author} {\bibinfo {author} {\bibfnamefont {A.}~\bibnamefont
  {Ariga}} \emph {et~al.} (\bibinfo {collaboration} {FASER}),\ }\href {\doibase
  10.1103/PhysRevD.99.095011} {\bibfield  {journal} {\bibinfo  {journal} {Phys.
  Rev. D}\ }\textbf {\bibinfo {volume} {99}},\ \bibinfo {pages} {095011}
  (\bibinfo {year} {2019})},\ \Eprint {http://arxiv.org/abs/1811.12522}
  {arXiv:1811.12522 [hep-ph]} \BibitemShut {NoStop}%
\bibitem [{\citenamefont {Delle~Rose}\ \emph {et~al.}(2019)\citenamefont
  {Delle~Rose}, \citenamefont {Khalil}, \citenamefont {King},\ and\
  \citenamefont {Moretti}}]{DelleRose:2018pgm}%
  \BibitemOpen
  \bibfield  {author} {\bibinfo {author} {\bibfnamefont {L.}~\bibnamefont
  {Delle~Rose}}, \bibinfo {author} {\bibfnamefont {S.}~\bibnamefont {Khalil}},
  \bibinfo {author} {\bibfnamefont {S.~J.~D.}\ \bibnamefont {King}}, \ and\
  \bibinfo {author} {\bibfnamefont {S.}~\bibnamefont {Moretti}},\ }\href
  {\doibase 10.3389/fphy.2019.00073} {\bibfield  {journal} {\bibinfo  {journal}
  {Front. in Phys.}\ }\textbf {\bibinfo {volume} {7}},\ \bibinfo {pages} {73}
  (\bibinfo {year} {2019})},\ \Eprint {http://arxiv.org/abs/1812.05497}
  {arXiv:1812.05497 [hep-ph]} \BibitemShut {NoStop}%
\bibitem [{\citenamefont {Hayes}\ \emph {et~al.}(2022)\citenamefont {Hayes},
  \citenamefont {Friar}, \citenamefont {Hale},\ and\ \citenamefont
  {Garvey}}]{Hayes:2021hin}%
  \BibitemOpen
  \bibfield  {author} {\bibinfo {author} {\bibfnamefont {A.~C.}\ \bibnamefont
  {Hayes}}, \bibinfo {author} {\bibfnamefont {J.~L.}\ \bibnamefont {Friar}},
  \bibinfo {author} {\bibfnamefont {G.~M.}\ \bibnamefont {Hale}}, \ and\
  \bibinfo {author} {\bibfnamefont {G.~T.}\ \bibnamefont {Garvey}},\ }\href
  {\doibase 10.1103/PhysRevC.105.055502} {\bibfield  {journal} {\bibinfo
  {journal} {Phys. Rev. C}\ }\textbf {\bibinfo {volume} {105}},\ \bibinfo
  {pages} {055502} (\bibinfo {year} {2022})},\ \Eprint
  {http://arxiv.org/abs/2106.06834} {arXiv:2106.06834 [nucl-th]} \BibitemShut
  {NoStop}%
\bibitem [{\citenamefont {Depero}\ \emph {et~al.}(2020)\citenamefont {Depero}
  \emph {et~al.}}]{NA64:2020xxh}%
  \BibitemOpen
  \bibfield  {author} {\bibinfo {author} {\bibfnamefont {E.}~\bibnamefont
  {Depero}} \emph {et~al.} (\bibinfo {collaboration} {NA64}),\ }\href {\doibase
  10.1140/epjc/s10052-020-08725-x} {\bibfield  {journal} {\bibinfo  {journal}
  {Eur. Phys. J. C}\ }\textbf {\bibinfo {volume} {80}},\ \bibinfo {pages}
  {1159} (\bibinfo {year} {2020})},\ \Eprint {http://arxiv.org/abs/2009.02756}
  {arXiv:2009.02756 [hep-ex]} \BibitemShut {NoStop}%
\bibitem [{\citenamefont {Seto}\ and\ \citenamefont
  {Shimomura}(2017)}]{Seto:2016pks}%
  \BibitemOpen
  \bibfield  {author} {\bibinfo {author} {\bibfnamefont {O.}~\bibnamefont
  {Seto}}\ and\ \bibinfo {author} {\bibfnamefont {T.}~\bibnamefont
  {Shimomura}},\ }\href {\doibase 10.1103/PhysRevD.95.095032} {\bibfield
  {journal} {\bibinfo  {journal} {Phys. Rev. D}\ }\textbf {\bibinfo {volume}
  {95}},\ \bibinfo {pages} {095032} (\bibinfo {year} {2017})},\ \Eprint
  {http://arxiv.org/abs/1610.08112} {arXiv:1610.08112 [hep-ph]} \BibitemShut
  {NoStop}%
\bibitem [{\citenamefont {Seto}\ and\ \citenamefont
  {Shimomura}(2021)}]{Seto:2020jal}%
  \BibitemOpen
  \bibfield  {author} {\bibinfo {author} {\bibfnamefont {O.}~\bibnamefont
  {Seto}}\ and\ \bibinfo {author} {\bibfnamefont {T.}~\bibnamefont
  {Shimomura}},\ }\href {\doibase 10.1007/JHEP04(2021)025} {\bibfield
  {journal} {\bibinfo  {journal} {JHEP}\ }\textbf {\bibinfo {volume} {04}},\
  \bibinfo {pages} {025} (\bibinfo {year} {2021})},\ \Eprint
  {http://arxiv.org/abs/2006.05497} {arXiv:2006.05497 [hep-ph]} \BibitemShut
  {NoStop}%
\bibitem [{\citenamefont {Ahmidouch}\ \emph {et~al.}(2021)\citenamefont
  {Ahmidouch} \emph {et~al.}}]{Ahmidouch:2021edo}%
  \BibitemOpen
  \bibfield  {author} {\bibinfo {author} {\bibfnamefont {A.}~\bibnamefont
  {Ahmidouch}} \emph {et~al.},\ }\href@noop {} {\  (\bibinfo {year} {2021})},\
  \Eprint {http://arxiv.org/abs/2108.13276} {arXiv:2108.13276 [nucl-ex]}
  \BibitemShut {NoStop}%
\bibitem [{\citenamefont {Rittenhouse~West}\ \emph {et~al.}(2021)\citenamefont
  {Rittenhouse~West}, \citenamefont {Brodsky}, \citenamefont {de~Teramond},
  \citenamefont {Goldhaber},\ and\ \citenamefont {Schmidt}}]{West:2020rlk}%
  \BibitemOpen
  \bibfield  {author} {\bibinfo {author} {\bibfnamefont {J.}~\bibnamefont
  {Rittenhouse~West}}, \bibinfo {author} {\bibfnamefont {S.~J.}\ \bibnamefont
  {Brodsky}}, \bibinfo {author} {\bibfnamefont {G.~F.}\ \bibnamefont
  {de~Teramond}}, \bibinfo {author} {\bibfnamefont {A.~S.}\ \bibnamefont
  {Goldhaber}}, \ and\ \bibinfo {author} {\bibfnamefont {I.}~\bibnamefont
  {Schmidt}},\ }\href {\doibase 10.1016/j.nuclphysa.2020.122134} {\bibfield
  {journal} {\bibinfo  {journal} {Nucl. Phys. A}\ }\textbf {\bibinfo {volume}
  {1007}},\ \bibinfo {pages} {122134} (\bibinfo {year} {2021})},\ \Eprint
  {http://arxiv.org/abs/2004.14659} {arXiv:2004.14659 [hep-ph]} \BibitemShut
  {NoStop}%
\bibitem [{\citenamefont {Dosch}\ \emph {et~al.}(2015)\citenamefont {Dosch},
  \citenamefont {de~Teramond},\ and\ \citenamefont {Brodsky}}]{Dosch:2015nwa}%
  \BibitemOpen
  \bibfield  {author} {\bibinfo {author} {\bibfnamefont {H.~G.}\ \bibnamefont
  {Dosch}}, \bibinfo {author} {\bibfnamefont {G.~F.}\ \bibnamefont
  {de~Teramond}}, \ and\ \bibinfo {author} {\bibfnamefont {S.~J.}\ \bibnamefont
  {Brodsky}},\ }\href {\doibase 10.1103/PhysRevD.91.085016} {\bibfield
  {journal} {\bibinfo  {journal} {Phys. Rev. D}\ }\textbf {\bibinfo {volume}
  {91}},\ \bibinfo {pages} {085016} (\bibinfo {year} {2015})},\ \Eprint
  {http://arxiv.org/abs/1501.00959} {arXiv:1501.00959 [hep-th]} \BibitemShut
  {NoStop}%
\bibitem [{\citenamefont {de~Teramond}\ \emph {et~al.}(2015)\citenamefont
  {de~Teramond}, \citenamefont {Dosch},\ and\ \citenamefont
  {Brodsky}}]{deTeramond:2014asa}%
  \BibitemOpen
  \bibfield  {author} {\bibinfo {author} {\bibfnamefont {G.~F.}\ \bibnamefont
  {de~Teramond}}, \bibinfo {author} {\bibfnamefont {H.~G.}\ \bibnamefont
  {Dosch}}, \ and\ \bibinfo {author} {\bibfnamefont {S.~J.}\ \bibnamefont
  {Brodsky}},\ }\href {\doibase 10.1103/PhysRevD.91.045040} {\bibfield
  {journal} {\bibinfo  {journal} {Phys. Rev. D}\ }\textbf {\bibinfo {volume}
  {91}},\ \bibinfo {pages} {045040} (\bibinfo {year} {2015})},\ \Eprint
  {http://arxiv.org/abs/1411.5243} {arXiv:1411.5243 [hep-ph]} \BibitemShut
  {NoStop}%
\bibitem [{\citenamefont {'t~Hooft}\ \emph {et~al.}(2008)\citenamefont
  {'t~Hooft}, \citenamefont {Isidori}, \citenamefont {Maiani}, \citenamefont
  {Polosa},\ and\ \citenamefont {Riquer}}]{tHooft:2008rus}%
  \BibitemOpen
  \bibfield  {author} {\bibinfo {author} {\bibfnamefont {G.}~\bibnamefont
  {'t~Hooft}}, \bibinfo {author} {\bibfnamefont {G.}~\bibnamefont {Isidori}},
  \bibinfo {author} {\bibfnamefont {L.}~\bibnamefont {Maiani}}, \bibinfo
  {author} {\bibfnamefont {A.~D.}\ \bibnamefont {Polosa}}, \ and\ \bibinfo
  {author} {\bibfnamefont {V.}~\bibnamefont {Riquer}},\ }\href {\doibase
  10.1016/j.physletb.2008.03.036} {\bibfield  {journal} {\bibinfo  {journal}
  {Phys. Lett. B}\ }\textbf {\bibinfo {volume} {662}},\ \bibinfo {pages} {424}
  (\bibinfo {year} {2008})},\ \Eprint {http://arxiv.org/abs/0801.2288}
  {arXiv:0801.2288 [hep-ph]} \BibitemShut {NoStop}%
\bibitem [{\citenamefont {Masjuan}\ and\ \citenamefont
  {Ruiz~Arriola}(2017)}]{Masjuan:2017fzu}%
  \BibitemOpen
  \bibfield  {author} {\bibinfo {author} {\bibfnamefont {P.}~\bibnamefont
  {Masjuan}}\ and\ \bibinfo {author} {\bibfnamefont {E.}~\bibnamefont
  {Ruiz~Arriola}},\ }\href {\doibase 10.1103/PhysRevD.96.054006} {\bibfield
  {journal} {\bibinfo  {journal} {Phys. Rev. D}\ }\textbf {\bibinfo {volume}
  {96}},\ \bibinfo {pages} {054006} (\bibinfo {year} {2017})},\ \Eprint
  {http://arxiv.org/abs/1707.05650} {arXiv:1707.05650 [hep-ph]} \BibitemShut
  {NoStop}%
\bibitem [{\citenamefont {Anselmino}\ \emph {et~al.}(1993)\citenamefont
  {Anselmino}, \citenamefont {Predazzi}, \citenamefont {Ekelin}, \citenamefont
  {Fredriksson},\ and\ \citenamefont {Lichtenberg}}]{Anselmino:1992vg}%
  \BibitemOpen
  \bibfield  {author} {\bibinfo {author} {\bibfnamefont {M.}~\bibnamefont
  {Anselmino}}, \bibinfo {author} {\bibfnamefont {E.}~\bibnamefont {Predazzi}},
  \bibinfo {author} {\bibfnamefont {S.}~\bibnamefont {Ekelin}}, \bibinfo
  {author} {\bibfnamefont {S.}~\bibnamefont {Fredriksson}}, \ and\ \bibinfo
  {author} {\bibfnamefont {D.~B.}\ \bibnamefont {Lichtenberg}},\ }\href
  {\doibase 10.1103/RevModPhys.65.1199} {\bibfield  {journal} {\bibinfo
  {journal} {Rev. Mod. Phys.}\ }\textbf {\bibinfo {volume} {65}},\ \bibinfo
  {pages} {1199} (\bibinfo {year} {1993})}\BibitemShut {NoStop}%
\bibitem [{\citenamefont {Nielsen}\ and\ \citenamefont
  {Brodsky}(2018)}]{Nielsen:2018uyn}%
  \BibitemOpen
  \bibfield  {author} {\bibinfo {author} {\bibfnamefont {M.}~\bibnamefont
  {Nielsen}}\ and\ \bibinfo {author} {\bibfnamefont {S.~J.}\ \bibnamefont
  {Brodsky}},\ }\href {\doibase 10.1103/PhysRevD.97.114001} {\bibfield
  {journal} {\bibinfo  {journal} {Phys. Rev. D}\ }\textbf {\bibinfo {volume}
  {97}},\ \bibinfo {pages} {114001} (\bibinfo {year} {2018})},\ \Eprint
  {http://arxiv.org/abs/1802.09652} {arXiv:1802.09652 [hep-ph]} \BibitemShut
  {NoStop}%
\bibitem [{\citenamefont {Brodsky}\ \emph {et~al.}(2015)\citenamefont
  {Brodsky}, \citenamefont {de~Teramond}, \citenamefont {Dosch},\ and\
  \citenamefont {Erlich}}]{Brodsky:2014yha}%
  \BibitemOpen
  \bibfield  {author} {\bibinfo {author} {\bibfnamefont {S.~J.}\ \bibnamefont
  {Brodsky}}, \bibinfo {author} {\bibfnamefont {G.~F.}\ \bibnamefont
  {de~Teramond}}, \bibinfo {author} {\bibfnamefont {H.~G.}\ \bibnamefont
  {Dosch}}, \ and\ \bibinfo {author} {\bibfnamefont {J.}~\bibnamefont
  {Erlich}},\ }\href {\doibase 10.1016/j.physrep.2015.05.001} {\bibfield
  {journal} {\bibinfo  {journal} {Phys. Rept.}\ }\textbf {\bibinfo {volume}
  {584}},\ \bibinfo {pages} {1} (\bibinfo {year} {2015})},\ \Eprint
  {http://arxiv.org/abs/1407.8131} {arXiv:1407.8131 [hep-ph]} \BibitemShut
  {NoStop}%
\bibitem [{\citenamefont {Brodsky}\ and\ \citenamefont
  {Chertok}(1976)}]{Brodsky:1976rz}%
  \BibitemOpen
  \bibfield  {author} {\bibinfo {author} {\bibfnamefont {S.~J.}\ \bibnamefont
  {Brodsky}}\ and\ \bibinfo {author} {\bibfnamefont {B.~T.}\ \bibnamefont
  {Chertok}},\ }\href {\doibase 10.1103/PhysRevD.14.3003} {\bibfield  {journal}
  {\bibinfo  {journal} {Phys. Rev. D}\ }\textbf {\bibinfo {volume} {14}},\
  \bibinfo {pages} {3003} (\bibinfo {year} {1976})}\BibitemShut {NoStop}%
\bibitem [{\citenamefont {Harvey}(1981{\natexlab{a}})}]{Harvey:1980rva}%
  \BibitemOpen
  \bibfield  {author} {\bibinfo {author} {\bibfnamefont {M.}~\bibnamefont
  {Harvey}},\ }\href {\doibase 10.1016/0375-9474(81)90413-9} {\bibfield
  {journal} {\bibinfo  {journal} {Nucl. Phys. A}\ }\textbf {\bibinfo {volume}
  {352}},\ \bibinfo {pages} {326} (\bibinfo {year}
  {1981}{\natexlab{a}})}\BibitemShut {NoStop}%
\bibitem [{\citenamefont {Brodsky}\ \emph {et~al.}(1983)\citenamefont
  {Brodsky}, \citenamefont {Ji},\ and\ \citenamefont
  {Lepage}}]{Brodsky:1983vf}%
  \BibitemOpen
  \bibfield  {author} {\bibinfo {author} {\bibfnamefont {S.~J.}\ \bibnamefont
  {Brodsky}}, \bibinfo {author} {\bibfnamefont {C.-R.}\ \bibnamefont {Ji}}, \
  and\ \bibinfo {author} {\bibfnamefont {G.~P.}\ \bibnamefont {Lepage}},\
  }\href {\doibase 10.1103/PhysRevLett.51.83} {\bibfield  {journal} {\bibinfo
  {journal} {Phys. Rev. Lett.}\ }\textbf {\bibinfo {volume} {51}},\ \bibinfo
  {pages} {83} (\bibinfo {year} {1983})}\BibitemShut {NoStop}%
\bibitem [{\citenamefont {Harvey}(1981{\natexlab{b}})}]{Harvey:1981udr}%
  \BibitemOpen
  \bibfield  {author} {\bibinfo {author} {\bibfnamefont {M.}~\bibnamefont
  {Harvey}},\ }\href {\doibase 10.1016/0375-9474(81)90412-7} {\bibfield
  {journal} {\bibinfo  {journal} {Nucl. Phys. A}\ }\textbf {\bibinfo {volume}
  {352}},\ \bibinfo {pages} {301} (\bibinfo {year} {1981}{\natexlab{b}})},\
  \bibinfo {note} {[Erratum: Nucl.Phys.A 481, 834 (1988)]}\BibitemShut
  {NoStop}%
\bibitem [{\citenamefont {Bashkanov}\ \emph {et~al.}(2013)\citenamefont
  {Bashkanov}, \citenamefont {Brodsky},\ and\ \citenamefont
  {Clement}}]{Bashkanov:2013cla}%
  \BibitemOpen
  \bibfield  {author} {\bibinfo {author} {\bibfnamefont {M.}~\bibnamefont
  {Bashkanov}}, \bibinfo {author} {\bibfnamefont {S.~J.}\ \bibnamefont
  {Brodsky}}, \ and\ \bibinfo {author} {\bibfnamefont {H.}~\bibnamefont
  {Clement}},\ }\href {\doibase 10.1016/j.physletb.2013.10.059} {\bibfield
  {journal} {\bibinfo  {journal} {Phys. Lett. B}\ }\textbf {\bibinfo {volume}
  {727}},\ \bibinfo {pages} {438} (\bibinfo {year} {2013})},\ \Eprint
  {http://arxiv.org/abs/1308.6404} {arXiv:1308.6404 [hep-ph]} \BibitemShut
  {NoStop}%
\bibitem [{\citenamefont {Miller}(2014)}]{Miller:2013hla}%
  \BibitemOpen
  \bibfield  {author} {\bibinfo {author} {\bibfnamefont {G.~A.}\ \bibnamefont
  {Miller}},\ }\href {\doibase 10.1103/PhysRevC.89.045203} {\bibfield
  {journal} {\bibinfo  {journal} {Phys. Rev. C}\ }\textbf {\bibinfo {volume}
  {89}},\ \bibinfo {pages} {045203} (\bibinfo {year} {2014})},\ \Eprint
  {http://arxiv.org/abs/1311.4561} {arXiv:1311.4561 [nucl-th]} \BibitemShut
  {NoStop}%
\bibitem [{\citenamefont {Rittenhouse~West}\ \emph {et~al.}(2020)\citenamefont
  {Rittenhouse~West}, \citenamefont {Brodsky}, \citenamefont {de~T\'eramond},\
  and\ \citenamefont {Schmidt}}]{RittenhouseWest:2019sar}%
  \BibitemOpen
  \bibfield  {author} {\bibinfo {author} {\bibfnamefont {J.}~\bibnamefont
  {Rittenhouse~West}}, \bibinfo {author} {\bibfnamefont {S.~J.}\ \bibnamefont
  {Brodsky}}, \bibinfo {author} {\bibfnamefont {G.~F.}\ \bibnamefont
  {de~T\'eramond}}, \ and\ \bibinfo {author} {\bibfnamefont {I.}~\bibnamefont
  {Schmidt}},\ }\href {\doibase 10.1016/j.physletb.2020.135423} {\bibfield
  {journal} {\bibinfo  {journal} {Phys. Lett. B}\ }\textbf {\bibinfo {volume}
  {805}},\ \bibinfo {pages} {135423} (\bibinfo {year} {2020})},\ \Eprint
  {http://arxiv.org/abs/1912.11288} {arXiv:1912.11288 [hep-ph]} \BibitemShut
  {NoStop}%
\bibitem [{\citenamefont {Jarmie}\ and\ \citenamefont
  {Allen}(1959)}]{Jarmie:1959zz}%
  \BibitemOpen
  \bibfield  {author} {\bibinfo {author} {\bibfnamefont {N.}~\bibnamefont
  {Jarmie}}\ and\ \bibinfo {author} {\bibfnamefont {R.~C.}\ \bibnamefont
  {Allen}},\ }\href {\doibase 10.1103/PhysRev.114.176} {\bibfield  {journal}
  {\bibinfo  {journal} {Phys. Rev.}\ }\textbf {\bibinfo {volume} {114}},\
  \bibinfo {pages} {176} (\bibinfo {year} {1959})}\BibitemShut {NoStop}%
\bibitem [{\citenamefont {Jarmie}\ \emph {et~al.}(1963)\citenamefont {Jarmie},
  \citenamefont {Silbert}, \citenamefont {Smith},\ and\ \citenamefont
  {Loos}}]{Jarmie:1963zz}%
  \BibitemOpen
  \bibfield  {author} {\bibinfo {author} {\bibfnamefont {N.}~\bibnamefont
  {Jarmie}}, \bibinfo {author} {\bibfnamefont {M.~G.}\ \bibnamefont {Silbert}},
  \bibinfo {author} {\bibfnamefont {D.~B.}\ \bibnamefont {Smith}}, \ and\
  \bibinfo {author} {\bibfnamefont {J.~S.}\ \bibnamefont {Loos}},\ }\href
  {\doibase 10.1103/PhysRev.130.1987} {\bibfield  {journal} {\bibinfo
  {journal} {Phys. Rev.}\ }\textbf {\bibinfo {volume} {130}},\ \bibinfo {pages}
  {1987} (\bibinfo {year} {1963})}\BibitemShut {NoStop}%
\bibitem [{\citenamefont {Walcher}(1970)}]{Walcher:1970vkv}%
  \BibitemOpen
  \bibfield  {author} {\bibinfo {author} {\bibfnamefont {T.}~\bibnamefont
  {Walcher}},\ }\href {\doibase 10.1016/0370-2693(70)90148-6} {\bibfield
  {journal} {\bibinfo  {journal} {Phys. Lett. B}\ }\textbf {\bibinfo {volume}
  {31}},\ \bibinfo {pages} {442} (\bibinfo {year} {1970})}\BibitemShut
  {NoStop}%
\bibitem [{\citenamefont {Jaffe}(2005)}]{Jaffe:2004ph}%
  \BibitemOpen
  \bibfield  {author} {\bibinfo {author} {\bibfnamefont {R.~L.}\ \bibnamefont
  {Jaffe}},\ }\href {\doibase 10.1016/j.physrep.2004.11.005} {\bibfield
  {journal} {\bibinfo  {journal} {Phys. Rept.}\ }\textbf {\bibinfo {volume}
  {409}},\ \bibinfo {pages} {1} (\bibinfo {year} {2005})},\ \Eprint
  {http://arxiv.org/abs/hep-ph/0409065} {arXiv:hep-ph/0409065} \BibitemShut
  {NoStop}%
\bibitem [{\citenamefont {West}(2023)}]{West:2020tyo}%
  \BibitemOpen
  \bibfield  {author} {\bibinfo {author} {\bibfnamefont {J.~R.}\ \bibnamefont
  {West}},\ }\href@noop {} {\bibfield  {journal} {\bibinfo  {journal}
  {Nuc.Phys.A}\ } (\bibinfo {year} {2023})},\ \Eprint
  {http://arxiv.org/abs/2009.06968} {arXiv:2009.06968 [hep-ph]} \BibitemShut
  {NoStop}%
\bibitem [{\citenamefont {Ilgenfritz}\ \emph {et~al.}(1978)\citenamefont
  {Ilgenfritz}, \citenamefont {Kripfganz},\ and\ \citenamefont
  {Schiller}}]{Ilgenfritz:1978cc}%
  \BibitemOpen
  \bibfield  {author} {\bibinfo {author} {\bibfnamefont {E.-M.}\ \bibnamefont
  {Ilgenfritz}}, \bibinfo {author} {\bibfnamefont {J.}~\bibnamefont
  {Kripfganz}}, \ and\ \bibinfo {author} {\bibfnamefont {A.}~\bibnamefont
  {Schiller}},\ }\href@noop {} {\bibfield  {journal} {\bibinfo  {journal} {Acta
  Phys. Polon. B}\ }\textbf {\bibinfo {volume} {9}},\ \bibinfo {pages} {881}
  (\bibinfo {year} {1978})}\BibitemShut {NoStop}%
\bibitem [{\citenamefont {Afanaciev}\ \emph {et~al.}(2024)\citenamefont
  {Afanaciev} \emph {et~al.}}]{MEGII:2024urz}%
  \BibitemOpen
  \bibfield  {author} {\bibinfo {author} {\bibfnamefont {K.}~\bibnamefont
  {Afanaciev}} \emph {et~al.} (\bibinfo {collaboration} {MEG II}),\ }\href@noop
  {} {\  (\bibinfo {year} {2024})},\ \Eprint {http://arxiv.org/abs/2411.07994}
  {arXiv:2411.07994 [nucl-ex]} \BibitemShut {NoStop}%
\bibitem [{\citenamefont {Batley}\ \emph {et~al.}(2015)\citenamefont {Batley}
  \emph {et~al.}}]{NA482:2015wmo}%
  \BibitemOpen
  \bibfield  {author} {\bibinfo {author} {\bibfnamefont {J.~R.}\ \bibnamefont
  {Batley}} \emph {et~al.} (\bibinfo {collaboration} {NA48/2}),\ }\href
  {\doibase 10.1016/j.physletb.2015.04.068} {\bibfield  {journal} {\bibinfo
  {journal} {Phys. Lett. B}\ }\textbf {\bibinfo {volume} {746}},\ \bibinfo
  {pages} {178} (\bibinfo {year} {2015})},\ \Eprint
  {http://arxiv.org/abs/1504.00607} {arXiv:1504.00607 [hep-ex]} \BibitemShut
  {NoStop}%
\bibitem [{\citenamefont {Banerjee}\ \emph {et~al.}(2020)\citenamefont
  {Banerjee} \emph {et~al.}}]{NA64:2019auh}%
  \BibitemOpen
  \bibfield  {author} {\bibinfo {author} {\bibfnamefont {D.}~\bibnamefont
  {Banerjee}} \emph {et~al.} (\bibinfo {collaboration} {NA64}),\ }\href
  {\doibase 10.1103/PhysRevD.101.071101} {\bibfield  {journal} {\bibinfo
  {journal} {Phys. Rev. D}\ }\textbf {\bibinfo {volume} {101}},\ \bibinfo
  {pages} {071101} (\bibinfo {year} {2020})},\ \Eprint
  {http://arxiv.org/abs/1912.11389} {arXiv:1912.11389 [hep-ex]} \BibitemShut
  {NoStop}%
\bibitem [{\citenamefont {Bossi}\ \emph {et~al.}(2023)\citenamefont {Bossi}
  \emph {et~al.}}]{PADME:2022tqr}%
  \BibitemOpen
  \bibfield  {author} {\bibinfo {author} {\bibfnamefont {F.}~\bibnamefont
  {Bossi}} \emph {et~al.} (\bibinfo {collaboration} {PADME}),\ }\href {\doibase
  10.1103/PhysRevD.107.012008} {\bibfield  {journal} {\bibinfo  {journal}
  {Phys. Rev. D}\ }\textbf {\bibinfo {volume} {107}},\ \bibinfo {pages}
  {012008} (\bibinfo {year} {2023})},\ \Eprint
  {http://arxiv.org/abs/2210.14603} {arXiv:2210.14603 [hep-ex]} \BibitemShut
  {NoStop}%
\bibitem [{\citenamefont {Darm\'e}\ \emph {et~al.}(2022)\citenamefont
  {Darm\'e}, \citenamefont {Mancini}, \citenamefont {Nardi},\ and\
  \citenamefont {Raggi}}]{Darme:2022zfw}%
  \BibitemOpen
  \bibfield  {author} {\bibinfo {author} {\bibfnamefont {L.}~\bibnamefont
  {Darm\'e}}, \bibinfo {author} {\bibfnamefont {M.}~\bibnamefont {Mancini}},
  \bibinfo {author} {\bibfnamefont {E.}~\bibnamefont {Nardi}}, \ and\ \bibinfo
  {author} {\bibfnamefont {M.}~\bibnamefont {Raggi}},\ }\href {\doibase
  10.1103/PhysRevD.106.115036} {\bibfield  {journal} {\bibinfo  {journal}
  {Phys. Rev. D}\ }\textbf {\bibinfo {volume} {106}},\ \bibinfo {pages}
  {115036} (\bibinfo {year} {2022})},\ \Eprint
  {http://arxiv.org/abs/2209.09261} {arXiv:2209.09261 [hep-ph]} \BibitemShut
  {NoStop}%
\end{thebibliography}%

\clearpage

\end{document}